\documentclass[sigconf]{acmart}
\usepackage{graphicx} 
\usepackage{url}
\usepackage{amsmath}
\usepackage{tabularx}
\usepackage{booktabs}
\usepackage{multirow}
\usepackage{threeparttable}
\usepackage{hyperref}

\AtBeginDocument{%
  }

\setcopyright{none}
\copyrightyear{2027}
\acmYear{2027}
\acmDOI{XXXXXXX.XXXXXXX}
\acmConference[WWW '27]
  {The ACM Web Conference 2027}
  {May 10--14, 2027}
  {Dublin, Ireland}
  
\begin{document}

\title{Role-Augmented Intent-Driven Generative Search Engine Optimization}

\author{Xiaolu Chen}
\orcid{0009-0001-9687-9923}
\affiliation{%
  \department{School of Cyber Science and Technology}
  \institution{University of Science and Technology of China}
  \city{Hefei}
  \state{Anhui}
  \country{China}
}
\email{xiaoluchen@mail.ustc.edu.cn}

\author{Jie Bao}
\orcid{0009-0001-5129-3945}
\affiliation{%
  \department{School of Cyber Science and Technology}
  \institution{University of Science and Technology of China}
  \city{Hefei}
  \state{Anhui}
  \country{China}
}
\email{baojie1996@mail.ustc.edu.cn}

\author{Haojie Wu}
\orcid{0009-0002-1848-4245}
\affiliation{%
  \department{School of Cyber Science and Technology}
  \institution{University of Science and Technology of China}
  \city{Hefei}
  \state{Anhui}
  \country{China}
}
\email{hjwuu@mail.ustc.edu.cn}

\author{Zhen Chen}
\correspondingauthor
\orcid{0009-0002-2151-7028}
\affiliation{%
  \department{School of Cyber Science and Technology}
  \institution{University of Science and Technology of China}
  \city{Hefei}
  \state{Anhui}
  \country{China}
}
\email{zhenchen@ustc.edu.cn}

\author{Liwei Pan}
\orcid{0009-0005-9935-0231}
\affiliation{%
  \department{School of Cyber Science and Technology}
  \institution{University of Science and Technology of China}
  \city{Hefei}
  \state{Anhui}
  \country{China}
}
\email{liweip@ustc.edu.cn}

\author{Yong Liao}
\orcid{0000-0001-6403-0557}
\affiliation{%
  \department{School of Cyber Science and Technology}
  \institution{University of Science and Technology of China}
  \city{Hefei}
  \state{Anhui}
  \country{China}
}
\email{yliao@ustc.edu.cn}

\renewcommand{\shortauthors}{Chen et al.}

\begin{abstract}
    Generative search engines (GSEs), which combine large language models (LLMs) with retrieval-augmented generation (RAG), are reshaping how users access Web information and making source visibility and influence in generated responses new optimization objectives. Existing generative search engine optimization (G-SEO) methods seek to improve content visibility, but many rely on direct access to target queries or GSE feedback elicited by those queries. However, Web content is typically optimized and published before future user queries are known, making G-SEO without access to target queries or their associated feedback an important yet underexplored problem.
    To address this, we propose RAID, a Role-Augmented Intent-Driven G-SEO method that operates without access to target queries or query-dependent GSE feedback. RAID infers a primary retrieval intent from source content, refines it through 4W multi-role deep reflection on prospective searchers' information needs, and uses both the primary and refined intents to jointly plan and rewrite source content. We further extend GEO-BENCH with diverse topic-related query variants, using both the original queries and their variants exclusively for evaluation to assess generalization to queries unseen during optimization. Across 100 samples and 500 evaluation queries, RAID achieves Relative Improvement Scores (RIS) of 3.00\% for Objective Overall and 1.32\% for Subjective Average, exceeding the best corresponding results among the nine predefined GEO rewriting strategies by 1.31 and 0.74 percentage points, respectively. With an optimization considered effective when $\mathrm{RIS}>0$, its effective optimization rates for Objective Overall and Subjective Average average 58.0\%, exceeding the highest corresponding average among the predefined strategies by 4.5 percentage points. These findings suggest that latent retrieval intent can serve as a semantic bridge between existing Web content and unknown future queries. Our code and extended dataset are publicly available at \url{https://anonymous.4open.science/r/RAID-G-SEO-F48F/}.
\end{abstract}



\received{XX XXXX XXXX}
\received[revised]{XX XXXX XXXX}
\received[accepted]{XX XXXX XXXX}

\maketitle

\section{Introduction}
Generative search engines (GSEs), which combine large language models (LLMs) with retrieval-augmented generation (RAG), are transforming how information is accessed and presented on the Web. As illustrated in Figure~\ref{fig:GSE_workflow_and_content_optimization_under_black-box,_query-unavailable_conditions}, traditional search engines primarily return ranked lists of links, whereas GSEs retrieve, select, and synthesize information from multiple Web sources to generate comprehensive responses to user queries. This paradigm shift makes the performance of Web content in generative search depend not only on retrieval rankings but also on whether its information is incorporated into generated responses, making source visibility and influence key optimization objectives.

\begin{figure*}[t]
\centering
\includegraphics[width=0.95\textwidth]{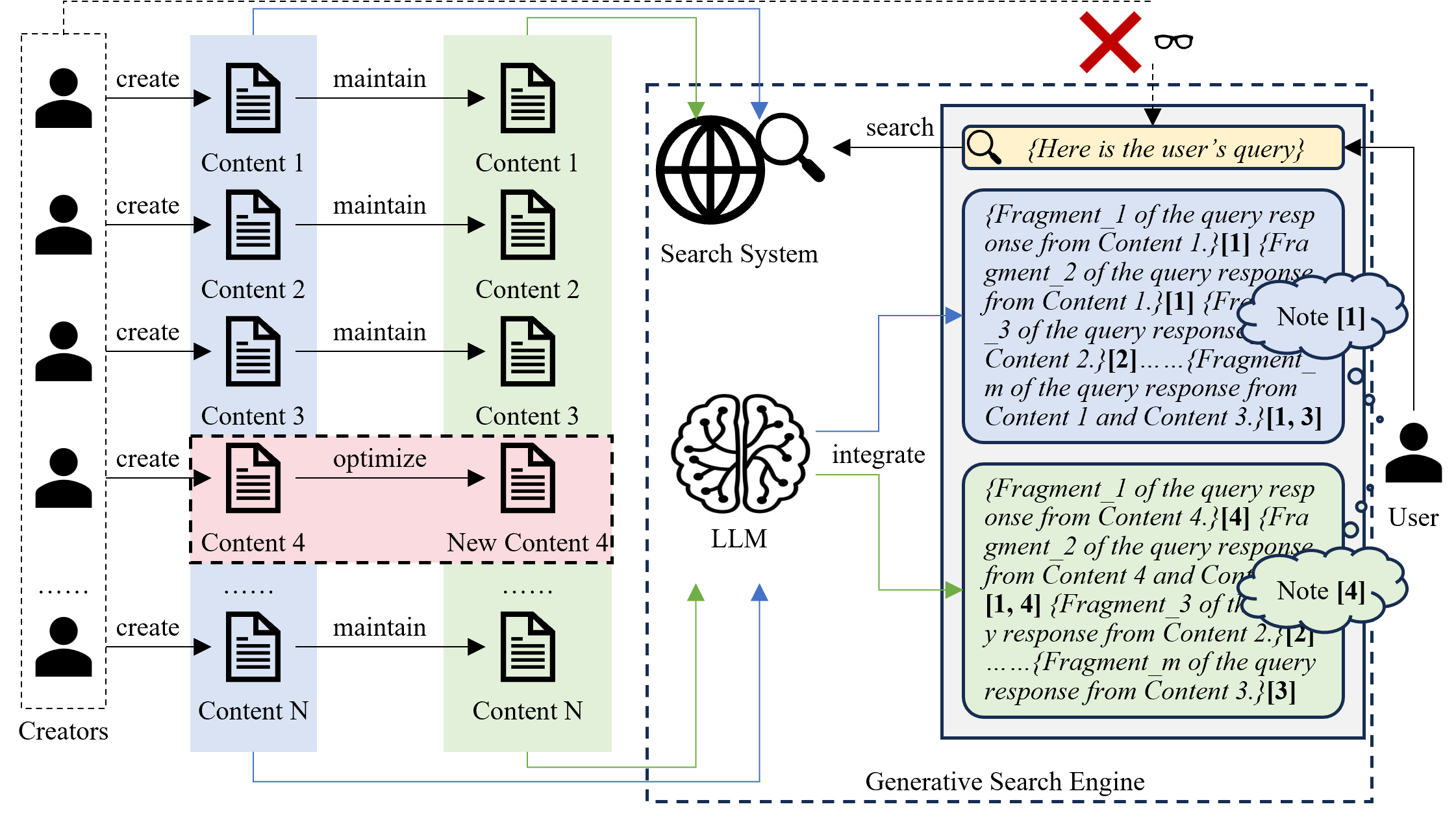} 
\caption{GSE workflow and content optimization under black-box, query-unavailable conditions. A GSE retrieves relevant sources and uses an LLM to generate a response with citations. Content optimizers can rewrite external sources (e.g., Content 4) without accessing target queries or query-dependent GSE feedback.}
\label{fig:GSE_workflow_and_content_optimization_under_black-box,_query-unavailable_conditions}
\end{figure*}

Traditional search engine optimization (SEO) primarily improves Web content rankings through techniques involving keywords, link structures, and page organization \cite{almukhtar2021searchengineoptimizationa}. Research on general-purpose text and prompt rewriting, including user-prompt reformulation \cite{sarkar2025conversationaluseraiinterventionstudy} and instruction-guided rewriting \cite{shu2024rewritelmaninstruction-tuned}, primarily focuses on rewriting quality or downstream task performance. However, neither line of research directly targets how information from Web sources is presented in generated responses. In contrast, generative search engine optimization (G-SEO) \cite{aggarwal2024geogenerativeengineoptimization} aims to improve source visibility. Existing approaches include prompt injection to influence product visibility and conversational search rankings \cite{kumar2024manipulatinglargelanguagemodels,pfrommer2024rankingmanipulationforconversational}, and source rewriting to improve the representation of target content in generated responses \cite{wu2026whatgenerativesearchengines,wu2026fromexperiencetoskill}.

However, many existing G-SEO methods still require access to target queries or rely on GSE feedback elicited by those queries to guide content optimization. Such dependencies are often difficult to satisfy when Web content must be optimized before future user queries become known. In practice, content optimizers may face two distinct constraints. One is \textit{system-side opacity}, as the mechanisms governing content retrieval, selection, and integration in black-box GSEs are inaccessible, making it difficult to determine how target content will be processed and presented. The other is \textit{query-side uncertainty}, since Web content is typically optimized and published before future user queries are known, preventing optimizers from accessing actual target queries or their corresponding GSE feedback in advance. Motivated by these constraints, we investigate \textit{Query-Unavailable G-SEO under Black-Box GSEs}, where optimizers can directly modify only Web source content without relying on internal GSE mechanisms, target queries, or query-dependent feedback. Under this setting, existing query-dependent methods are difficult to apply directly. This raises the question of \textit{how Web content can be effectively optimized for generative search using only existing source content when GSEs are black boxes and target queries are unavailable}.

Although target queries are unavailable, the topics and semantics of Web content may still reveal the latent information needs of prospective users, providing a basis for inferring retrieval intent. However, the same source may serve different searcher roles whose needs vary in focus, required detail, and presentation preferences. Consequently, a single inferred intent may not adequately capture these heterogeneous needs. Motivated by this observation, we propose RAID, the Role-Augmented Intent-Driven Generative Search Engine Optimization method. RAID treats latent retrieval intent as a semantic bridge between existing Web content and unknown future queries and follows four stages comprising content summarization, intent inference and refinement, optimization planning, and content rewriting. Specifically, RAID infers a primary retrieval intent from source content and its summary, refines it through 4W (Who, What, Why, and How) multi-role deep reflection on complementary information needs of prospective searcher roles while instructing the model to ground the refinement in existing source information, and uses both the primary and refined intents to guide optimization planning and content rewriting. In this way, RAID derives optimization guidance from existing content to accommodate diverse potential retrieval needs without relying on specific target queries or query-dependent feedback.

To evaluate RAID under query-unavailable conditions, we extend GEO-BENCH \cite{aggarwal2024geogenerativeengineoptimization} with diverse topic-related query variants reflecting different potential information needs. Both the original queries and their variants are used exclusively for evaluation and do not participate in the content optimization process. Experiments in a simulated GSE setting with fixed source sets show that RAID improves source visibility without access to target queries or query-dependent GSE feedback, with average gains observed across diverse unseen query variants. Across 100 samples and 500 evaluation queries, RAID achieves Relative Improvement Scores (RIS) of 3.00\% for Objective Overall and 1.32\% for Subjective Average, exceeding the best corresponding results among the nine predefined GEO rewriting strategies by 1.31 and 0.74 percentage points, respectively. RAID also achieves higher mean RIS than all nine strategies in each of the seven subjective evaluation dimensions. With effective optimization defined as RIS greater than zero, RAID achieves objective and subjective effective optimization rates of 57.0\% and 59.0\%, respectively, exceeding the highest corresponding rates among the predefined strategies by 2.8 and 6.2 percentage points. The arithmetic mean of these rates reaches 58.0\%. These results demonstrate the effectiveness of RAID's source-side retrieval intent modeling approach for query-unavailable G-SEO. Since RAID's public release, subsequent studies have explored intent-driven G-SEO from different perspectives \cite{zhou2026ifgeoconflictawareinstructionfusion,chen2026mindreaderlatentuser,ramakrishna2026queryimpliedgenerativeengine}, reflecting the continued development of this research direction.


Our main contributions are summarized below:
\begin{itemize}
    \item To the best of our knowledge, we are the first to explicitly formulate and systematically investigate query-unavailable G-SEO under black-box GSEs, characterizing the dual constraints of system-side opacity and query-side uncertainty.
    \item We propose RAID, a Role-Augmented Intent-Driven G-SEO method that refines source-inferred intent through 4W analysis of complementary searcher needs and jointly plans and rewrites content under query-unavailable conditions.
    \item We extend GEO-BENCH with diverse topic-related query variants, using both the original queries and their variants exclusively for evaluation, and demonstrate RAID's effectiveness in improving source visibility on unseen queries in a simulated GSE setting.
\end{itemize}

\section{Related Work}
\subsection{Search Engine Optimization}
Traditional SEO seeks to improve webpage rankings through content optimization \cite{iqbal2022searchengineoptimizationseo}, keyword selection \cite{nagpal2021keywordselectionstrategiesin,kanara2024pythondrivenkeywordanalysis}, link building \cite{shahzad2020thenewtrendfor,almukhtar2021searchengineoptimizationa}, and technical optimization \cite{kowalczyk2024EnhancingSEOinsingle-page}. These techniques primarily target conventional search mechanisms \cite{lewandowski2023understandingearchengines}. Recent studies have leveraged LLMs for SEO-oriented content generation \cite{chodak2024largelanguagemodelsfor,samarah2024utilizingllmsforenhancing} and ranking-oriented document editing \cite{bardas2025automaticdocumenteditingfor}. However, these approaches primarily focus on retrieval rankings rather than source visibility in generative search responses.

\subsection{Text Rewriting}
Text rewriting involves modifying textual expressions while preserving their original semantics. RewriteLM \cite{shu2024rewritelmaninstruction-tuned} employs instruction tuning for general-purpose text rewriting, while multi-intent text revision \cite{chong2023leveragingprefixtransferfor} uses editing intents to guide modifications. Other studies have explored prompt rewriting for personalized or conversational generation \cite{li2024learningtorewriteprompts,sarkar2025conversationaluseraiinterventionstudy}, contextual search intent modeling \cite{maoetal2023largelanguagemodelsknow}, as well as user intent alignment in domain-specific settings \cite{li2025deepthinkaligninglanguagemodels}. These studies explore text rewriting and intent alignment across different generation and retrieval tasks, but do not directly optimize source visibility in generative search responses.

\subsection{Generative Search Engine Optimization}
G-SEO focuses on enhancing source visibility in responses generated by GSEs. Related security research has demonstrated that textual interventions can influence the behavior of LLM-based systems \cite{greshake2023notwhatyouvesigned,shi2024optimizationbasedpromptinjection, chen2026unveilingtheresilienceoflLMenhanced}. In generative search scenarios, adversarial studies have further explored product visibility manipulation \cite{kumar2024manipulatinglargelanguagemodels,nestaas2025adversarialsearchengineoptimization} and prompt-injection-based search ranking manipulation \cite{pfrommer2024rankingmanipulationforconversational}. Meanwhile, GEO \cite{aggarwal2024geogenerativeengineoptimization} systematically formulated the problem of source visibility optimization and introduced GEO-BENCH, establishing a foundation for subsequent G-SEO research and evaluation.

In recent G-SEO research, AutoGEO \cite{wu2026whatgenerativesearchengines} learns content preferences of GSEs, MAGEO \cite{wu2026fromexperiencetoskill} distills editing experiences into reusable optimization strategies, and FeatGEO \cite{liu2026thinkbeforewritingfeaturelevel} optimizes interpretable document features. Despite employing different optimization mechanisms, these approaches rely to varying degrees either directly on target queries or indirectly on query-triggered GSE feedback for optimization guidance, which may limit their direct applicability when neither target queries nor their associated feedback are available. RAID instead operates under a source-only optimization setting in which target queries and their associated GSE feedback are unavailable.

Furthermore, G-SEO evaluation benchmarks have expanded to conversational optimization \cite{Puerto2025cseobenchdoesconversational}, end-to-end retrieval and generation \cite{kim2026sageoarenaarealistic}, source influence assessment \cite{chen2025ccgseobenchcontentcentricbenchmarkmeasuring}, and e-commerce scenarios \cite{bagga2026egeotestbedgenerativeengine}. Although these benchmarks broaden the evaluation scope across different tasks and system stages, they differ in their assumptions about query availability during optimization. To evaluate generalization under a strict query-unavailable setting, we further extend GEO-BENCH with topic-related query variants, ensuring that neither the original queries, their variants, nor the associated GSE feedback is used during content optimization.

Since RAID was publicly released, subsequent studies have further explored intent-driven G-SEO approaches. IF-GEO \cite{zhou2026ifgeoconflictawareinstructionfusion} adopts RAID's extended evaluation benchmark and coordinates conflicting revision requirements across latent queries. Mind Reader \cite{chen2026mindreaderlatentuser} includes RAID as an experimental baseline and extracts implicit information needs from given queries. QI-GEO \cite{ramakrishna2026queryimpliedgenerativeengine} also discusses RAID and identifies relevant information gaps through document semantic relations and knowledge graph expansion. These studies further advance intent-driven G-SEO from different perspectives.

\begin{figure*}[t]
\centering
\includegraphics[width=0.95\textwidth]{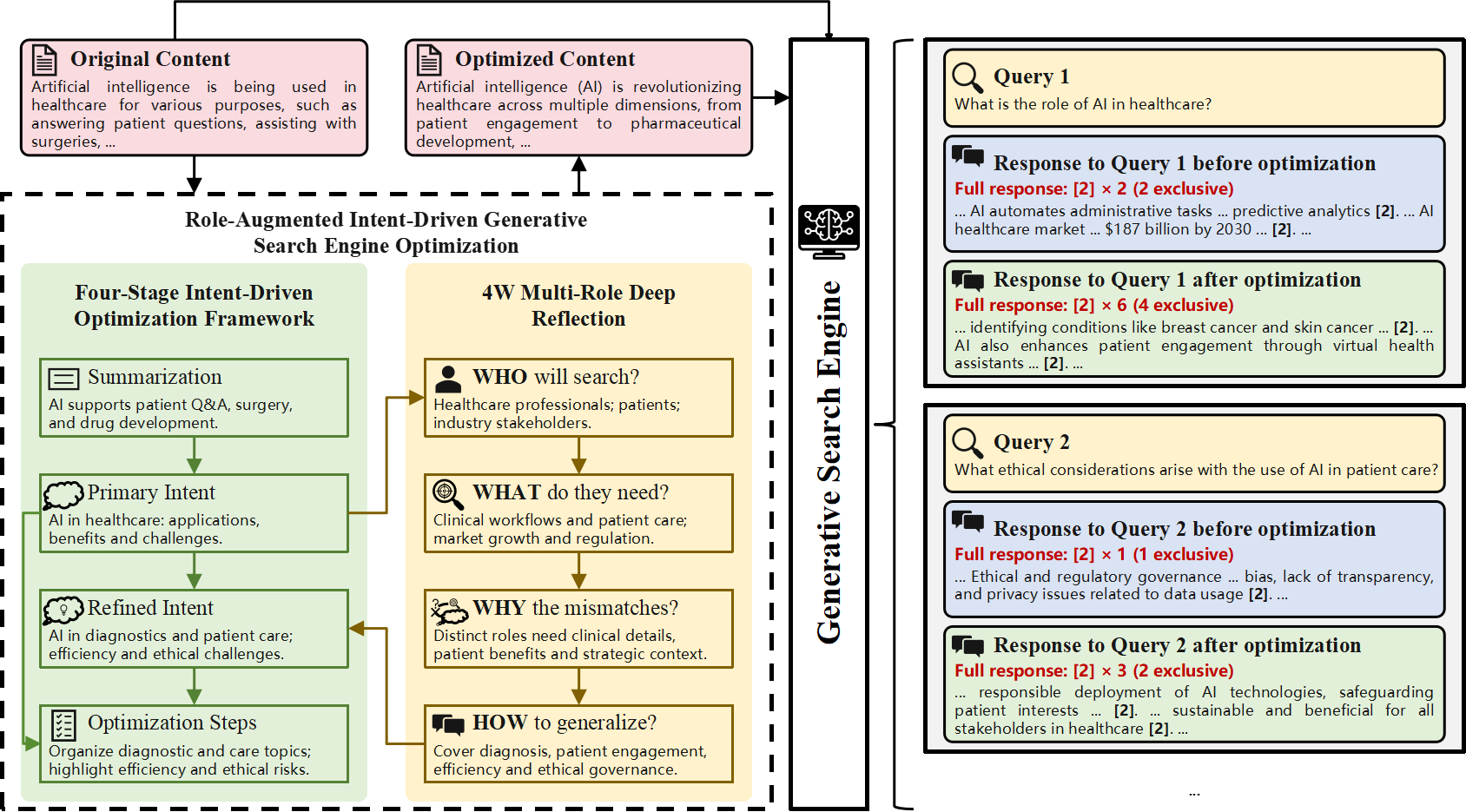} 
\caption{Role-Augmented Intent-Driven G-SEO framework. RAID summarizes source content, infers a primary search intent and refines it through 4W multi-role deep reflection, plans optimization steps, and rewrites the content without accessing target queries or query-dependent GSE feedback. The AI-in-healthcare example compares the target source's citation and exclusive-citation counts in the full responses to two queries before and after rewriting.}
\label{fig:Role-Augmented_Intent-Driven_G-SEO_framework}
\end{figure*}

\section{Methodology}
\subsection{Query-Unavailable G-SEO under Black-Box GSEs}
We follow the black-box GSE setting adopted by GEO \cite{aggarwal2024geogenerativeengineoptimization}, as illustrated in Figure~\ref{fig:GSE_workflow_and_content_optimization_under_black-box,_query-unavailable_conditions}. Given a user query $q$, a GSE first retrieves a set of relevant Web content sources and employs an internal LLM to generate a response conditioned on the query and retrieved content:
\begin{equation}
\begin{aligned}
C(q) &= \operatorname{Retrieval}(q)
= \{c_1,c_2,\ldots,c_N\},\\
r(q) &= \operatorname{Generate}(q,C(q)).
\end{aligned}
\end{equation}
Here, $C(q)$ denotes the set of retrieved sources, and $r(q)$ represents the generated response. The response can be represented as an ordered sequence of sentences, $r(q)=(l_1,l_2,\ldots,l_m)$. For each sentence $l_j$ containing source citations, let $C_j\subseteq C(q)$ denote the set of sources cited by that sentence. These citations and the presentation of source content in the response provide signals for assessing source visibility.
Under this black-box GSE setting, optimizers cannot access or modify the internal retrieval and generation mechanisms and can influence system outputs only by modifying external content sources. We further consider the query-unavailable constraint, under which target queries and their associated GSE feedback are inaccessible during content optimization. Specifically, given a target content source $c$, the optimizer produces a rewritten version $c'$ using only the source and information derived from it, aiming to improve its visibility in GSE responses to future relevant queries. This optimization process is expressed as:
\begin{equation}
c' = \operatorname{RAID}(c).
\end{equation}
The central challenge is therefore to derive optimization guidance from source content that can accommodate diverse potential information needs in the absence of query-side signals. Directly rewriting the original content may reinforce information already emphasized in the source while overlooking other relevant user needs. To address this challenge, we propose RAID, a Role-Augmented Intent-Driven optimization method that infers a latent search intent from source content, refines it through structured multi-role information need analysis, and combines the primary and refined intents to guide content modification, aiming to balance optimization specificity with applicability across unseen queries.

\subsection{Four-Stage Intent-Driven Optimization Framework}
We define search intent as the information need or objective that a potential user seeks to satisfy through information retrieval. Under the query-unavailable setting, the optimizer cannot directly observe intents expressed in target queries and must instead infer potential information needs from source content. RAID therefore employs inferred search intent as semantic guidance connecting potential user needs with source content and organizes optimization into four stages, as illustrated in Figure~\ref{fig:Role-Augmented_Intent-Driven_G-SEO_framework}.
\begin{itemize}
    \item \textbf{Step 1: Content Summarization.}
    Given a target content source $c$, RAID first generates a semantic summary $s$ that captures its central topic, key information, and main points. This stage aims to highlight the source's essential semantic information while reducing interference from redundant expressions and secondary details. The summary complements the original content for subsequent intent inference.
    \item \textbf{Step 2: Intent Inference and Refinement.}
    Based on the original content $c$ and its summary $s$, RAID infers a primary retrieval intent $i_0$ to characterize potential information needs associated with the source. However, a single retrieval intent may not adequately capture the heterogeneous needs of different prospective searcher roles. RAID therefore introduces 4W multi-role deep reflection to examine complementary information needs from different role perspectives and refine the primary intent into $i^*$ based on information available in the original source.
    \item \textbf{Step 3: Step Planning.}
    Based on the primary intent $i_0$ and the refined intent $i^*$, RAID develops a content optimization strategy. This stage prioritizes the primary intent while incorporating complementary information needs reflected in the refined intent. It identifies relevant information that may be insufficiently emphasized in the original source and plans how to organize and present it to improve coverage of potential retrieval needs.
    \item \textbf{Step 4: Content Rewriting.}
    RAID rewrites the original content $c$ according to the optimization strategy, producing the optimized content $c'$. This stage aims to preserve the source's core meaning and key information while improving the organization and presentation of relevant information to better address the needs of different prospective searcher roles, thereby enhancing source visibility in generative search responses.
\end{itemize}
The four-stage process can be summarized conceptually as:
\begin{equation}
\begin{aligned}
s &= \operatorname{Summarize}(c),\\
(i_0,\mathcal{R},i^*) &= \operatorname{InferRefine}_{\mathrm{4W}}(c,s),\\
c' &= \operatorname{PlanAndRewrite}(c,s,i_0,i^*).
\end{aligned}
\end{equation}
Here, $s$ denotes the source summary, $i_0$ the primary search intent, $\mathcal{R}$ the multi-role 4W analysis, and $i^*$ the refined search intent. The optimization strategy is not produced as a separate intermediate output, and the entire process operates without target queries or query-dependent GSE feedback.

\subsection{4W Multi-Role Deep Reflection}
Inferring search intents from source content presents two related challenges. First, an initially inferred intent may primarily reflect information already expressed in the source, potentially overlooking the heterogeneous needs of different users. Second, unconstrained intent expansion may introduce information needs that are only weakly related to the original content, reducing the specificity of optimization guidance. Intent refinement must therefore broaden the range of potential needs considered while preserving semantic relevance to the original content and primary intent. Prior studies have investigated reflection and multi-perspective reasoning to improve LLM inference \cite{shinn2023reflexionlanguageagentswith,ji2023towardsmitigatingllmhallucination,zhangetal2024selfcontrastbetterreflectionthrough,yanetal2024mirrormultipleperspectiveselfreflectionmethod}. Inspired by structured question-based analysis in interdisciplinary research \cite{Ward2017WhywhosewhatandhowAframeworkforknowledgemobilisers,Škėrienė2020problemsolvingthroughvalues}, we design a 4W multi-role deep reflection mechanism tailored to search intent refinement. Organized around \textit{Who}, \textit{What}, \textit{Why}, and \textit{How}, the mechanism analyzes potentially overlooked needs through four interconnected components covering user role identification, information need analysis, discrepancy identification, and intent refinement. It uses these insights to refine the primary intent while maintaining its core semantic focus.
\begin{itemize}
    \item \textbf{Who: Who might search for this content?}
    Given the primary search intent $i_0$, the model infers representative user roles that may have information needs related to the source. Different roles may have distinct occupational backgrounds, levels of expertise, and areas of interest, leading to different information needs. Grounding role identification in the primary intent introduces diverse yet relevant user perspectives without arbitrary role expansion.
    \item \textbf{What: What information do these roles seek?}
    For each prospective searcher role, the model identifies the core information that the role would seek from the original source. This component translates role perspectives into concrete information needs, providing a basis for subsequent comparison with the primary intent and intent refinement.
    \item \textbf{Why: Why might the initial intent not fully address these needs?}
    The model examines which role-specific information needs remain implicit in the primary intent or determines whether the primary intent already addresses them sufficiently. This component identifies potentially overlooked needs without assuming that every role requires additional information.
    \item \textbf{How: How should the initial intent be refined?}
    Based on the preceding analysis, the model refines the intent from each role's perspective to make relevant information needs explicit. The primary need and useful complementary role-specific needs are then integrated into a coherent refined intent $i^*$, restricted to information supported by the original source. If no useful refinement emerges, the primary intent is retained.
\end{itemize}
Through this structured reflection process, RAID identifies information needs that remain implicit in the primary intent from different prospective searcher perspectives and integrates useful complementary needs into a source-grounded refined intent. The primary and refined intents jointly guide subsequent content optimization, with the refined intent providing complementary coverage requirements rather than replacing the primary intent. A structured overview of RAID's core prompts is provided in Appendix~\ref{app:Core Prompts Used in RAID G-SEO}.

\section{Experiments}
\subsection{Experimental Setup}
The software and hardware configurations used in our experiments are detailed in Appendix~\ref{app:Software and Hardware Environment}.

\subsubsection{Generative Search Engine Simulation}
We follow the GSE simulation framework of GEO \cite{aggarwal2024geogenerativeengineoptimization}, using the pre-retrieved source sets provided by GEO-BENCH and modeling response generation as a single-turn task. Specifically, for each evaluated query, we use the fixed set of five content sources associated with its GEO-BENCH sample as the generation context and employ GLM-4-9B-0414 to generate the response. For each query, we generate five responses under each of the original and rewritten source conditions. The model's generation parameters are detailed in Appendix~\ref{app:Configuration Parameters of LLMs}.

\begin{table}[t]
\centering
\caption{An original GEO-BENCH query and its four topic-related variants reflecting distinct information needs.}
\label{tab:An_original_GEO-BENCH_query_and_its_four_topic-related_variants}
\footnotesize
\renewcommand{\arraystretch}{1.08}
\begin{tabularx}{\columnwidth}{@{}>{\raggedright\arraybackslash}X@{}}
\toprule
\textbf{How does quantum computing work?}\\
1. What are the practical applications of quantum computing?\par
2. What are the limitations and challenges of quantum computing?\par
3. How does quantum computing differ from classical computing?\par
4. How can beginners use Qiskit to experiment with quantum computing?\\
\bottomrule
\end{tabularx}
\end{table}

\begin{table*}[t]
\centering
\caption{Objective and subjective performance of G-SEO methods on the expanded GEO-BENCH (100 samples). Objective Overall and Subjective Average serve as the primary comparison metrics, with other columns providing fine-grained results. Higher RIS (\%) indicates better performance. Boldface and underlining mark the best and second-best results per metric. $\dagger$ indicates a significant difference from RAID (two-sided paired Wilcoxon signed-rank test, Holm-adjusted $p<0.05$ across 10 comparisons per composite metric).}
\label{tab:Objective_and_subjective_performance_of_G-SEO_methods_on_the_expanded_GEO-BENCH__100_samples_}

\footnotesize
\renewcommand{\arraystretch}{1.05}
\setlength{\tabcolsep}{4pt}

\begin{tabular*}{\textwidth}{@{\extracolsep{\fill}}l*{11}{r}@{}}
\toprule

& \multicolumn{3}{c}{\textbf{Objective RIS}}
& \multicolumn{8}{c}{\textbf{Subjective RIS}} \\

\cmidrule(lr){2-4}
\cmidrule(l){5-12}

\textbf{Method}
& Word Count
& Posi. Count
& \textbf{Overall}
& Rele.
& Inf.
& Uniq.
& Dive.
& Clic.
& Sub. Posi.
& Sub. Volu.
& \textbf{Average} \\

\midrule

\multicolumn{12}{@{}l}{\textit{Predefined Rewriting Strategies (GEO)}} \\
\addlinespace[2pt]

Tran. SEO
& 0.44 & 0.23 & 0.22\textsuperscript{$\dagger$}
& 0.10 & -0.01 & 0.26 & -0.04
& -0.04 & 0.12 & 0.16 & 0.02\textsuperscript{$\dagger$} \\

Uniq. Word
& -1.44 & -1.45 & -1.50\textsuperscript{$\dagger$}
& -0.60 & -0.43 & -1.98 & -0.38
& -0.74 & -0.53 & -0.97 & -0.82\textsuperscript{$\dagger$} \\

Simp. Expr.
& -0.89 & -1.05 & -1.05\textsuperscript{$\dagger$}
& 0.06 & 0.10 & -0.51 & -0.08
& -0.04 & 0.19 & -0.03 & -0.08\textsuperscript{$\dagger$} \\

Auth. Expr.
& 1.23 & 1.17 & 1.16
& 0.55 & 0.33 & 1.36 & 0.25
& 0.78 & 0.64 & 0.79 & 0.58 \\

Flue. Expr.
& 1.08 & 1.08 & 0.98
& 0.28 & 0.12 & 1.27 & 0.11
& 0.30 & 0.22 & 0.30 & 0.27\textsuperscript{$\dagger$} \\

Term. Addi.
& 1.38 & 1.38 & 1.36
& 0.21 & 0.29 & 1.28 & 0.08
& 0.12 & 0.30 & 0.64 & 0.33 \\

Repu. Addi.
& 1.48 & 1.31 & 1.28
& 0.31 & 0.14 & 1.11 & 0.02
& 0.48 & 0.21 & 0.24 & 0.28\textsuperscript{$\dagger$} \\

Quot. Addi.
& 1.78 & 1.68 & 1.69
& 0.45 & 0.24 & 1.36 & 0.16
& 0.51 & 0.52 & 0.64 & 0.47 \\

Stat. Addi.
& 1.26 & 1.33 & 1.24
& 0.49 & 0.33 & 1.55 & 0.17
& 0.64 & 0.50 & 0.52 & 0.50\textsuperscript{$\dagger$} \\

\midrule

\multicolumn{12}{@{}l}{\textit{Intent-Driven Optimization Methods}} \\
\addlinespace[2pt]

\textbf{RAID}
& \underline{3.06} & \underline{3.04} & \underline{3.00}
& \underline{1.05} & \underline{0.83}
& \underline{3.92} & \underline{0.55}
& \underline{1.46} & \underline{1.20}
& \underline{1.55} & \underline{1.32} \\

IF-GEO
& \textbf{5.56} & \textbf{5.40} & \textbf{5.35}\textsuperscript{$\dagger$}
& \textbf{1.39} & \textbf{1.02}
& \textbf{5.30} & \textbf{0.58}
& \textbf{1.67} & \textbf{1.47}
& \textbf{2.31} & \textbf{1.75} \\

\bottomrule
\end{tabular*}
\end{table*}

\subsubsection{Dataset}
We extend GEO-BENCH \cite{aggarwal2024geogenerativeengineoptimization} to evaluate the generalization of G-SEO methods to diverse potential information needs under the query-unavailable setting. GEO-BENCH is a comprehensive benchmark for G-SEO evaluation, comprising queries collected from real-world search systems, complex reasoning and debate-oriented questions, and synthetic queries generated by LLMs. Each original query is associated with five relevant content sources. To simulate heterogeneous information needs related to the same topic, we use GPT-4 to generate four topic-related variants for each of the 1,000 original queries in the GEO-BENCH test split. These variants explore the original topic from different perspectives, representing diverse potential information needs rather than semantically equivalent paraphrases. Each extended sample consists of one original query, four query variants, and the five content sources associated with the original query, with all queries in the group sharing the same fixed source set. Neither the original queries nor their variants are exposed during content optimization. Instead, they are used for performance evaluation after optimization, without providing the optimizer with target-query-triggered GSE feedback. To ensure data quality, we conducted multiple rounds of validation and revision, focusing on the topical relevance of query variants and whether the fixed source sets substantively support their core information needs. After excluding three anomalous groups with pre-existing mismatches between the original queries and their associated source sets, we retained 997 valid extended samples. Details of query generation and quality validation are provided in Appendix~\ref{app:Query Variant Generation and Validation}, and illustrative examples are presented in Table~\ref{tab:An_original_GEO-BENCH_query_and_its_four_topic-related_variants}. We randomly select 100 samples from the 997 retained groups for experimental evaluation to assess the performance of content optimization methods across diverse potential information needs.

\subsubsection{Baselines}
To maintain consistency in optimization tasks and input conditions, we adopt the nine source-only content rewriting implementations released with GEO \cite{aggarwal2024geogenerativeengineoptimization} as our primary baselines. These implementations take only the source content to be optimized as task-specific input, covering different approaches to stylistic modification and content enrichment. We additionally include IF-GEO \cite{zhou2026ifgeoconflictawareinstructionfusion} as another representative intent-driven baseline. As a subsequent study inspired by RAID's search intent modeling approach, IF-GEO guides content rewriting by mining latent queries and coordinating their optimization requirements, enabling a comparison of different approaches to modeling and leveraging latent search intents. All methods operate under the query-unavailable constraint, without access to the target queries used for evaluation or their associated GSE feedback during content optimization. We implement all optimization methods using GLM-4-9B-0414 and evaluate them under the same protocol to reduce the influence of model differences on the comparison.

\subsubsection{Evaluation Metrics}
We evaluate content optimization effectiveness using the objective visibility and subjective impression metrics proposed by GEO \cite{aggarwal2024geogenerativeengineoptimization}. The objective metric, Position-Adjusted Word Count (PAWC), considers the number and positions of words attributed to a source in generated responses, assigning higher visibility scores to sources cited earlier and more extensively. The subjective metric, Subjective Impression, comprises seven dimensions: relevance, influence, diversity, uniqueness, follow-up likelihood, subjective position, and subjective count. For subjective evaluation, we adopt a stepwise scoring approach adapted from G-Eval \cite{liuetal2023gevalnlgevalutationusing}. To clarify the evaluation criteria for each dimension, we adopt a prompt-generate-prompt approach, using GPT-4o to generate dimension-specific scoring criteria and evaluation steps with a unified 0--5 scoring scale. The actual scoring is performed by GLM-4-9B-0414. The scoring criteria generation prompt and the unified evaluation template are provided in Appendix~\ref{app:G-eval Prompt Generation Prompt and Evaluation Prompt Template}. We measure content optimization effectiveness using the Relative Improvement Score (RIS), which compares source visibility scores before and after optimization. For a target source $C_i$ and query $q$, let $r_q$ and $r'_q$ denote the generated responses before and after optimization, respectively. The RIS is defined as:
\begin{equation}
    \mathrm{RIS}_{C_i}^{(m)}(q)
    =
    \frac{
        \mathrm{Imp}_{C_i}^{(m)}(r'_q)
        -
        \mathrm{Imp}_{C_i}^{(m)}(r_q)
    }{
        \mathrm{Imp}_{C_i}^{(m)}(r_q)+1
    }
    \times 100\%,
    \label{eq:Relative Improvement Score}
\end{equation}
where $\mathrm{Imp}_{C_i}^{(m)}$ denotes the visibility score of source $C_i$ under metric $m$, representing an objective or subjective evaluation metric. We introduce a smoothing constant of 1 in the denominator to avoid division by zero when the original visibility score is zero. For the aggregate subjective metric Average, we first compute the arithmetic mean of the raw scores across the seven subjective dimensions and then calculate the corresponding RIS, rather than averaging the dimension-wise RIS values. For each extended sample, we calculate the RIS for the original query and its four variants and average them to obtain the sample-level result for each metric. Statistical inference is conducted at the extended-sample level (100 samples), using paired Wilcoxon signed-rank tests with Holm correction and 95\% confidence intervals from 20,000 bootstrap resamples. To validate LLM-based subjective scoring, one human evaluator rated target sources in responses to 250 queries from 50 samples using the same seven dimensions and 0--5 scale as the model. Across the seven dimensions, human and LLM scores achieved a mean Spearman rank correlation of 0.69 and a mean quadratic-weighted Cohen's kappa of 0.57, indicating positive ranking association and moderate agreement.

\begin{figure*}[t]
\centering
\includegraphics[width=0.95\textwidth]{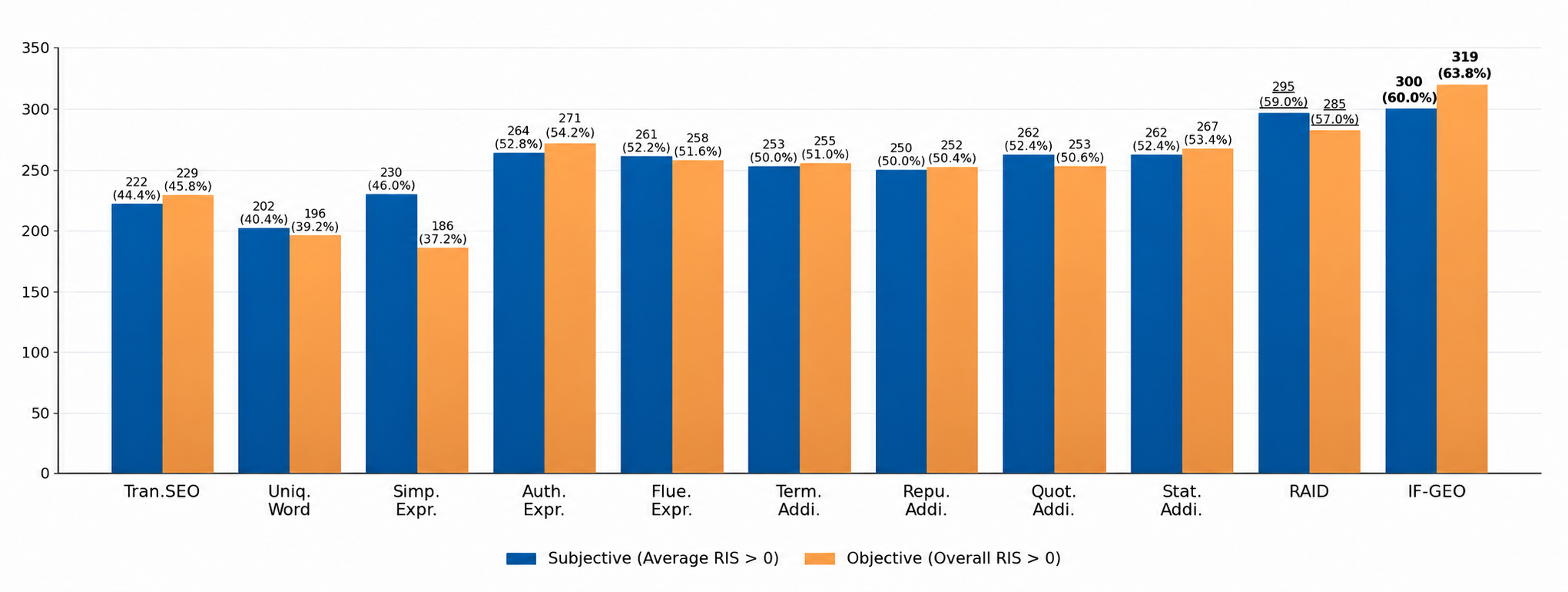}
\caption{Adaptability of G-SEO methods on the expanded GEO-BENCH (100 samples). Across 500 query-level instances, blue and orange bars show effective optimization counts (RIS $>0$) for Subjective Average and Objective Overall, respectively. Labels indicate counts and rates, with boldface and underlining marking the best and second-best results.}
\label{fig:Adaptability_analysis_of_G-SEO_methods_on_the_expanded_GEO-BENCH__100_samples_}
\end{figure*}

\subsection{Results and Analysis}
\subsubsection{Main Results}
Table~\ref{tab:Objective_and_subjective_performance_of_G-SEO_methods_on_the_expanded_GEO-BENCH__100_samples_} presents the objective and subjective visibility performance of all methods on the expanded GEO-BENCH. RAID achieves higher mean RIS than all nine predefined GEO rewriting strategies across all 11 objective and subjective evaluation metrics. Specifically, RAID achieves an Objective Overall RIS of 3.00\% (95\% CI [1.49\%, 4.62\%]) and a Subjective Average RIS of 1.32\% (95\% CI [0.64\%, 2.05\%]), exceeding the best corresponding results among these strategies by 1.31 and 0.74 percentage points, respectively. After Holm correction, RAID significantly outperforms three and six of the predefined strategies in Objective Overall and Subjective Average, respectively. These results demonstrate that source-side latent search intent can effectively guide content optimization and yield positive gains in both objective and subjective visibility without access to target queries or query-dependent GSE feedback. We further compare RAID with IF-GEO \cite{zhou2026ifgeoconflictawareinstructionfusion}, a subsequent intent-driven method. IF-GEO achieves an Objective Overall RIS of 5.35\% and a Subjective Average RIS of 1.75\%. The difference is statistically significant for Objective Overall after Holm correction but not for Subjective Average. The two methods differ in how latent search intents are modeled and utilized. RAID refines a primary search intent inferred from source content through multi-role information need analysis, whereas IF-GEO explicitly mines multiple latent queries and coordinates their optimization requirements through priority-driven conflict resolution. Overall, RAID demonstrates that refining source-inferred retrieval intent through the complementary information needs of prospective searcher roles can improve objective and subjective source visibility under query-unavailable conditions, highlighting the value of intent-driven G-SEO.

\subsubsection{Adaptability Analysis}
Average RIS summarizes mean optimization performance but does not indicate how frequently positive gains occur across queries. To assess the cross-query adaptability of G-SEO methods under query-unavailable conditions, we examine 500 query-level instances from the expanded GEO-BENCH. For each evaluation dimension, an optimization is considered effective when its corresponding RIS exceeds zero. We report the counts and rates separately for Objective Overall and Subjective Average. As shown in Figure~\ref{fig:Adaptability_analysis_of_G-SEO_methods_on_the_expanded_GEO-BENCH__100_samples_}, RAID achieves 295 subjective (59.0\%) and 285 objective (57.0\%) effective optimizations, exceeding Auth. Expr., the best-performing predefined GEO strategy in the corresponding dimensions (52.8\% and 54.2\%), by 6.2 and 2.8 percentage points, respectively. The arithmetic mean of these rates reaches 58.0\%, exceeding Auth. Expr. (53.5\%) by 4.5 percentage points. These results show that RAID produces positive gains on more than half of the evaluated queries under both metrics, complementing its advantages in mean RIS. IF-GEO achieves the highest subjective and objective effective optimization rates among all methods, at 60.0\% and 63.8\%, respectively. Notably, both effective optimization rates remain below 70\% across all evaluated methods, indicating further opportunities to improve G-SEO generalization across diverse unseen queries.

\begin{table}[t]
\centering
\caption{Performance comparison of RAID and its four ablation variants on the expanded GEO-BENCH (100 samples), reporting mean Objective Overall and Subjective Average RIS.}
\label{tab:Performance_comparison_of_RAID_and_its_ablation_variants_on_the_expanded_GEO-BENCH__100_samples_}

\footnotesize
\renewcommand{\arraystretch}{1.05}
\setlength{\tabcolsep}{2pt}

\begin{tabular*}{\columnwidth}{@{\extracolsep{\fill}}lcc@{}}
\toprule
\textbf{Method} & \textbf{Objective Overall} & \textbf{Subjective Average} \\
\midrule
RAID & \underline{3.00} & \textbf{1.32} \\
w/o 4W Refl. & 2.14 & \underline{1.21} \\
w/o Summ. & \textbf{3.16} & 1.13 \\
w/o Inte. & 1.62 & 1.05 \\
w/o Plan. & 2.66 & 0.74 \\
\bottomrule
\end{tabular*}
\end{table}

\begin{figure}[t]
\centering
\includegraphics[width=0.95\columnwidth]{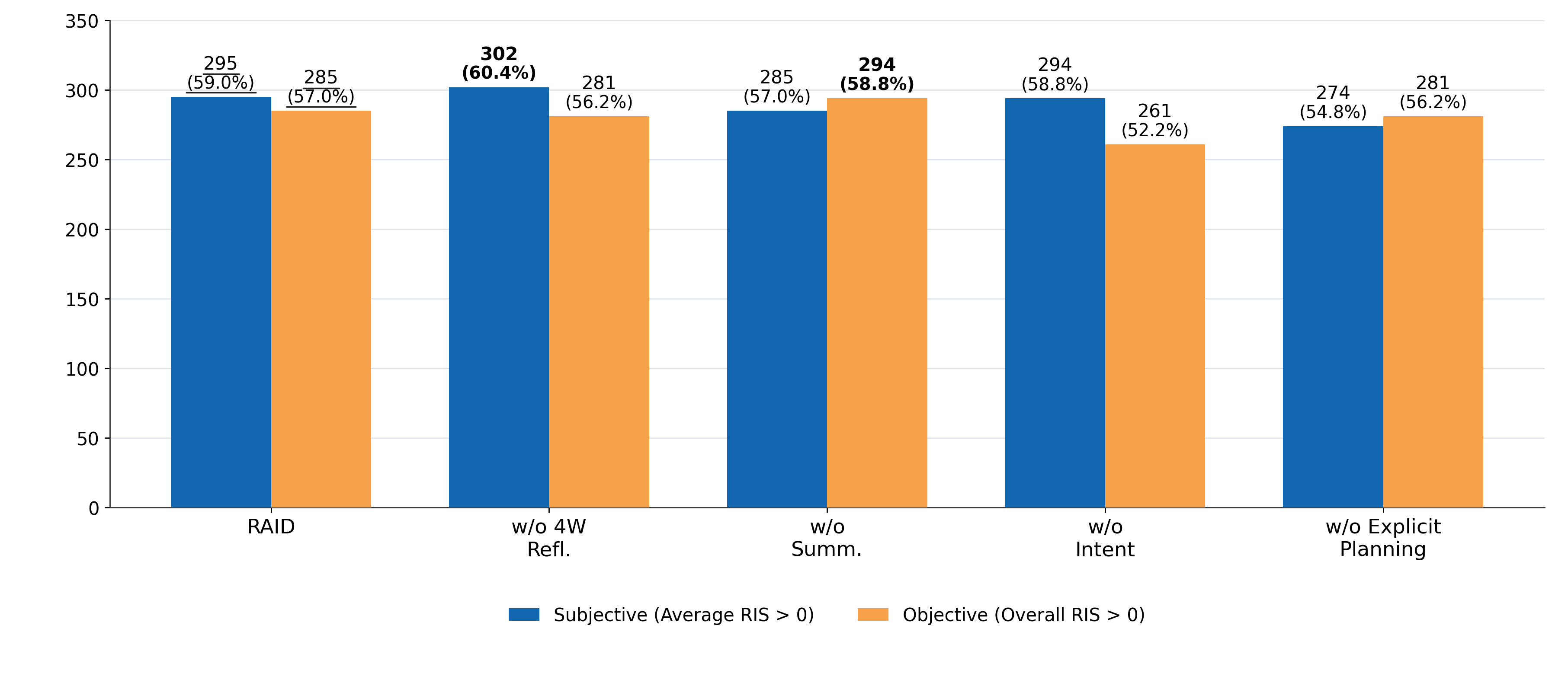}
\caption{Adaptability comparison of RAID and its ablation variants on the expanded GEO-BENCH (100 samples). Bars show effective optimization counts and rates (RIS $>0$).}
\label{fig:Adaptability_comparison_of_RAID_and_its_ablation_variants_on_the_expanded_GEO-BENCH__100_samples_}
\end{figure}

\begin{figure}[t]
\centering
\includegraphics[width=0.95\columnwidth]{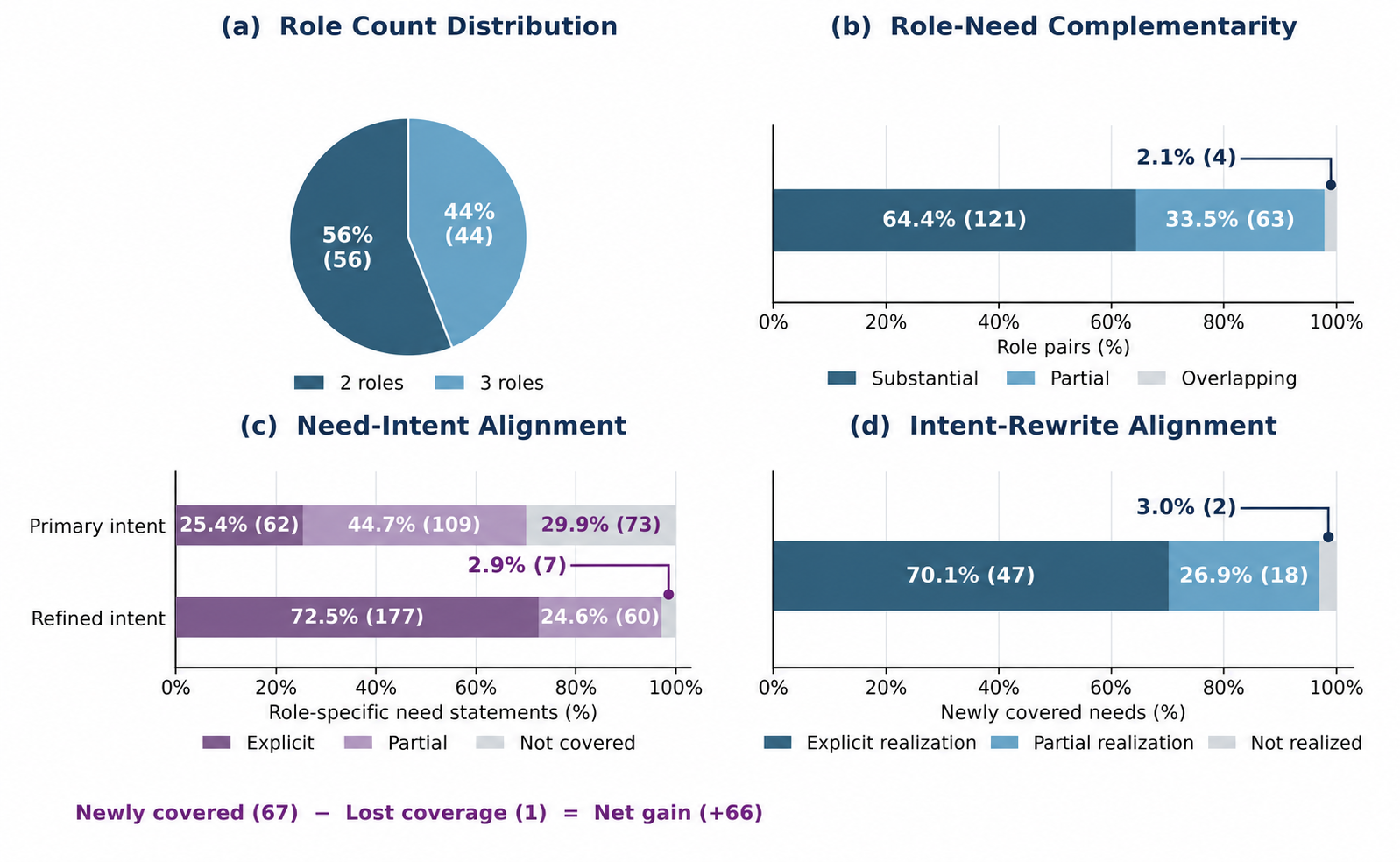}
\caption{Optimization behavior analysis of RAID. (a) Distribution of prospective searcher roles across 100 sources. (b) Information-need complementarity across 188 within-source role pairs. (c) Semantic coverage of 244 role-specific need statements in primary and refined intents. (d) Realization of 67 newly covered needs in the final source rewrites.}
\label{fig:Optimization_behavior_analysis_of_RAID}
\end{figure}

\subsubsection{Ablation Study}
To assess the roles of RAID's intent-guided workflow and its key components, we evaluate four matched ablation variants that remove 4W multi-role deep reflection, content summarization, all intent inference and refinement, or explicit optimization planning instructions. In the variant without 4W reflection, relevant information directly derived from the original source replaces multi-role reflection to produce the refined intent. All configurations follow the query-unavailable setting and use the same sources, evaluation queries, and target sources. Table~\ref{tab:Performance_comparison_of_RAID_and_its_ablation_variants_on_the_expanded_GEO-BENCH__100_samples_} and Figure~\ref{fig:Adaptability_comparison_of_RAID_and_its_ablation_variants_on_the_expanded_GEO-BENCH__100_samples_} report composite RIS and effective optimization rates, respectively. The results show that RAID achieves the highest Subjective Average RIS and the second-highest Objective Overall RIS, while ranking second in both subjective and objective effective optimization rates. It is the only configuration ranking among the top two across all four metrics, demonstrating a relatively balanced performance across objective and subjective visibility gains and effective optimization rates. In comparison, removing summarization improves objective performance but reduces subjective performance, whereas replacing 4W reflection yields the highest subjective effective optimization rate but a lower objective RIS than full RAID. Removing all intent inference and refinement results in a significant reduction in Objective Overall RIS (Holm-adjusted $p=0.0275$). Overall, these findings support the contribution of inferred intent information to objective visibility and reveal metric-dependent effects of the individual components.

\subsubsection{Optimization Behavior Analysis}
We further examine RAID's optimization behavior through role count distribution, role–need complementarity, need–intent alignment, and intent–rewrite alignment. Role–need complementarity is assessed at the within-source role-pair level, distinguishing no, partial, and substantial complementarity. The two alignment analyses use each role's WHAT statement as a role-specific need unit and distinguish no, partial, and explicit semantic correspondence. The semantic annotations were manually reviewed. As shown in Figure~\ref{fig:Optimization_behavior_analysis_of_RAID}(a) and (b), RAID identifies two prospective searcher roles for 56 of the 100 sources and three roles for the remaining 44, yielding 244 role-specific need statements and 188 within-source role pairs. Among these pairs, 121 exhibit substantial complementarity and 63 show partial complementarity, giving a combined rate of 97.87\%. Figure~\ref{fig:Optimization_behavior_analysis_of_RAID}(c) shows that partial or explicit coverage of role-specific needs increases from 171 of 244 (70.08\%) in the primary intents to 237 of 244 (97.13\%) in the refined intents, a gain of 27.05 percentage points with a source-cluster bootstrap 95\% confidence interval of [21.81, 32.41]. As shown in Figure~\ref{fig:Optimization_behavior_analysis_of_RAID}(d), among the 67 role-specific needs covered by refined intents but not primary intents, 47 are explicitly reflected in the final rewritten content and 18 are partially reflected, yielding a rate of at least partial realization of 97.01\%. These findings trace multi-role information needs through intent refinement to observable content-level outcomes under query-unavailable optimization.

\subsubsection{Case Study}
We examine a gift-recommendation source from the expanded GEO-BENCH to illustrate how a single RAID rewrite performs across unseen queries. Detailed source revisions and before-and-after GSE response excerpts are provided in Appendix~\ref{app:A Case of RAID G-SEO}. Without access to the original query (Q0) or its four topic-related variants (Q1--Q4), RAID rewrites only the target source [3], making existing gift categories explicitly address strangers and reorganizing gift suggestions and explanations while reducing the source from 1,437 to 875 whitespace-delimited tokens. For the query about safe and neutral birthday gifts (Q1), objective Overall RIS and subjective Average RIS reach 3.21 and 12.71, respectively, while citations to [3] increase from one to two in the illustrated response. For the query about unique but universally appealing gifts (Q3), the two RIS values reach 5.42 and 12.06, with citations to [3] increasing from zero to two. By comparison, citations to [3] decrease from five to one for the original query (Q0), with an objective RIS of -2.57 but a subjective RIS of 23.15. For another query (Q4), citations increase from two to four, with an objective RIS of 7.04 but a subjective RIS of -1.35. These observations show that citation count changes in individual responses do not necessarily correspond to objective and subjective visibility gains aggregated over multiple generations. Overall, the case illustrates how a single query-unavailable rewrite can achieve positive gains across multiple related unseen queries with varying outcomes across queries and evaluation metrics.

\section{Conclusion}
This work is the first to explicitly formulate query-unavailable G-SEO under black-box GSEs and proposes RAID, a role-augmented intent-driven content optimization method. RAID employs a four-stage optimization framework to infer a primary search intent from source content, refine it through 4W multi-role deep reflection, and use the primary and refined intents to plan and guide content rewriting. To evaluate generalization to unseen queries, we further extend GEO-BENCH with topic-related query variants, using both original queries and their variants exclusively for evaluation. Experimental results demonstrate that RAID achieves improvements in both objective and subjective source visibility on unseen queries, validating the effectiveness of latent search intent as semantic guidance for query-unavailable G-SEO.

\bibliographystyle{ACM-Reference-Format}
\bibliography{main/references}

@inproceedings{aggarwal2024geogenerativeengineoptimization,
author = {Aggarwal, Pranjal and Murahari, Vishvak and Rajpurohit, Tanmay and Kalyan, Ashwin and Narasimhan, Karthik and Deshpande, Ameet},
title = {GEO: Generative Engine Optimization},
year = {2024},
isbn = {9798400704901},
publisher = {Association for Computing Machinery},
address = {New York, NY, USA},
url = {https://doi.org/10.1145/3637528.3671900},
doi = {10.1145/3637528.3671900},
booktitle = {Proceedings of the 30th ACM SIGKDD Conference on Knowledge Discovery and Data Mining},
pages = {5–16},
numpages = {12},
location = {Barcelona, Spain},
series = {KDD '24}
}

@inproceedings{zhou2026ifgeoconflictawareinstructionfusion,
    title = "{IF}-{GEO}: Conflict-Aware Instruction Fusion for Multi-Query Generative Engine Optimization",
    author = "Zhou, Heyang  and
      Chen, Jiajia  and
      Chen, Xiaolu  and
      Bao, Jie  and
      Chen, Zhen  and
      Liao, Yong",
    editor = "Liakata, Maria  and
      Moreira, Viviane P.  and
      Zhang, Jiajun  and
      Jurgens, David",
    booktitle = "Findings of the {A}ssociation for {C}omputational {L}inguistics: {ACL} 2026",
    month = jul,
    year = "2026",
    address = "San Diego, California, United States",
    publisher = "Association for Computational Linguistics",
    url = "https://aclanthology.org/2026.findings-acl.1373/",
    doi = "10.18653/v1/2026.findings-acl.1373",
    pages = "27576--27590",
    ISBN = "979-8-89176-395-1"
}

@inproceedings{kim2026sageoarenaarealistic,
author = {Kim, Sunghwan and Jeong, Wooseok and Kim, Serin and Lee, Sangam and Lee, Dongha},
title = {SAGEO Arena: A Realistic Environment for Evaluating Search-Augmented Generative Engine Optimization},
year = {2026},
isbn = {9798400722592},
publisher = {Association for Computing Machinery},
address = {New York, NY, USA},
url = {https://doi.org/10.1145/3770855.3818146},
doi = {10.1145/3770855.3818146},
booktitle = {Proceedings of the 32nd ACM SIGKDD Conference on Knowledge Discovery and Data Mining V.2},
pages = {2342–2353},
numpages = {12},
location = {Republic of Korea},
series = {KDD '26}
}

@inproceedings{wu2026fromexperiencetoskill,
    title = "From Experience to Skill: Multi-Agent Generative Engine Optimization via Reusable Strategy Learning",
    author = "Wu, Beining  and
      Mao, Fuyou  and
      Lin, Jiong  and
      Yang, Cheng  and
      Lu, Jiaxuan  and
      Guo, Yifu  and
      Zhang, Siyu  and
      Wu, Yifan  and
      Huang, Ying  and
      Li, Fu",
    editor = "Liakata, Maria  and
      Moreira, Viviane P.  and
      Zhang, Jiajun  and
      Jurgens, David",
    booktitle = "Findings of the {A}ssociation for {C}omputational {L}inguistics: {ACL} 2026",
    month = jul,
    year = "2026",
    address = "San Diego, California, United States",
    publisher = "Association for Computational Linguistics",
    url = "https://aclanthology.org/2026.findings-acl.2149/",
    doi = "10.18653/v1/2026.findings-acl.2149",
    pages = "43305--43315",
    ISBN = "979-8-89176-395-1"
}

@inproceedings{chen2026mindreaderlatentuser,
    title = "Mind Reader: Latent User Demand-Guided Content Optimization for Generative Search Engine",
    author = "Chen, Tong  and
      Guo, JiaWei  and
      Li, Yuxi  and
      Chen, Baiming  and
      Ren, Houxing  and
      Zhiwei, Zhang  and
      Zhang, Yunxiang  and
      Xia, Hanyang  and
      Liang, Kun  and
      Fan, Zhaoran",
    editor = "Liakata, Maria  and
      Moreira, Viviane P.  and
      Zhang, Jiajun  and
      Jurgens, David",
    booktitle = "Proceedings of the 64th Annual Meeting of the {A}ssociation for {C}omputational {L}inguistics (Volume 1: Long Papers)",
    month = jul,
    year = "2026",
    address = "San Diego, California, United States",
    publisher = "Association for Computational Linguistics",
    url = "https://aclanthology.org/2026.acl-long.1894/",
    doi = "10.18653/v1/2026.acl-long.1894",
    pages = "40832--40848",
    ISBN = "979-8-89176-390-6"
}

@inproceedings{wu2026whatgenerativesearchengines,
 author = {Wu, Yujiang and Zhong, Shanshan and Kim, Yubin and Xiong, Chenyan},
 booktitle = {International Conference on Learning Representations},
 editor = {C. Vondrick and B. Hariharan and C. Raffel and L. Pinto and D. Yang and A. Faust},
 pages = {136858--136888},
 title = {What Generative Search Engines Like and How to Optimize Web Content Cooperatively},
 url = {https://proceedings.iclr.cc/paper_files/paper/2026/file/dd5dfba659a7ec010414de1c1debdeb4-Paper-Conference.pdf},
 volume = {2026},
 year = {2026}
}

@inproceedings{Puerto2025cseobenchdoesconversational,
 author = {Puerto, Haritz and Gubri, Martin and Green, Tommaso and Oh, Seong Joon and Yun, Sangdoo},
 booktitle = {Advances in Neural Information Processing Systems},
 doi = {10.52202/085713-0923},
 editor = {D. Belgrave and C. Zhang and H. Lin and R. Pascanu and P. Koniusz and M. Ghassemi and N. Chen},
 pages = {},
 publisher = {Curran Associates, Inc.},
 title = {C-SEO Bench: Does Conversational SEO Work?},
 url = {https://proceedings.neurips.cc/paper_files/paper/2025/file/27aa3aeff0f8460a7b43d30fa6c5c032-Paper-Datasets_and_Benchmarks_Track.pdf},
 volume = {38, Main Conference},
 year = {2025}
}

@inproceedings{liu2026thinkbeforewritingfeaturelevel,
    title = "Think Before Writing: Feature-Level Multi-Objective Optimization for Generative Citation Visibility",
    author = "Liu, Zikang  and
      Xu, Peilan",
    editor = "Liakata, Maria  and
      Moreira, Viviane P.  and
      Zhang, Jiajun  and
      Jurgens, David",
    booktitle = "Proceedings of the 64th Annual Meeting of the {A}ssociation for {C}omputational {L}inguistics (Volume 1: Long Papers)",
    month = jul,
    year = "2026",
    address = "San Diego, California, United States",
    publisher = "Association for Computational Linguistics",
    url = "https://aclanthology.org/2026.acl-long.929/",
    doi = "10.18653/v1/2026.acl-long.929",
    pages = "20290--20303",
    ISBN = "979-8-89176-390-6"
}

@inproceedings{chen2026unveilingtheresilienceoflLMenhanced,
author = {Chen, Pei and Hong, Geng and Wu, Xinyi and Wu, Mengying and Zhu, Zixuan and Liu, Mingxuan and Liu, Baojun and Zhang, Mi and Yang, Min},
title = {Unveiling the Resilience of LLM-Enhanced Search Engines against Black-Hat SEO Manipulation},
year = {2026},
isbn = {9798400723070},
publisher = {Association for Computing Machinery},
address = {New York, NY, USA},
url = {https://doi.org/10.1145/3774904.3792399},
doi = {10.1145/3774904.3792399},
booktitle = {Proceedings of the ACM Web Conference 2026},
pages = {2962–2973},
numpages = {12},
location = {United Arab Emirates},
series = {WWW '26}
}

@article{almukhtar2021searchengineoptimizationa,
  title={Search engine optimization: a review},
  author={Almukhtar, Firas and Mahmoodd, Nawzad and Kareem, Shahab},
  journal={Applied computer science},
  volume={17},
  number={1},
  pages={70--80},
  year={2021},
  publisher={Polskie Towarzystwo Promocji Wiedzy}
}

@InProceedings{chodak2024largelanguagemodelsfor,
author="Chodak, Grzegorz
and B{\l}a{\.{z}}yczek, Klaudia",
editor="Garg, Deepak
and Rodrigues, Joel J. P. C.
and Gupta, Suneet Kumar
and Cheng, Xiaochun
and Sarao, Pushpender
and Patel, Govind Singh",
title="Large Language Models for Search Engine Optimization in E-commerce",
booktitle="Advanced Computing",
year="2024",
publisher="Springer Nature Switzerland",
address="Cham",
pages="333--344",
isbn="978-3-031-56700-1"
}

@inproceedings{chong2023leveragingprefixtransferfor,
    title = "Leveraging Prefix Transfer for Multi-Intent Text Revision",
    author = "Chong, Ruining  and
      Kong, Cunliang  and
      Wu, Liu  and
      Liu, Zhenghao  and
      Jin, Ziye  and
      Yang, Liner  and
      Fan, Yange  and
      Fan, Hanghang  and
      Yang, Erhong",
    editor = "Rogers, Anna  and
      Boyd-Graber, Jordan  and
      Okazaki, Naoaki",
    booktitle = "Proceedings of the 61st Annual Meeting of the Association for Computational Linguistics (Volume 2: Short Papers)",
    month = jul,
    year = "2023",
    address = "Toronto, Canada",
    publisher = "Association for Computational Linguistics",
    url = "https://aclanthology.org/2023.acl-short.105/",
    doi = "10.18653/v1/2023.acl-short.105",
    pages = "1219--1228"
}

@inproceedings{greshake2023notwhatyouvesigned,
author = {Greshake, Kai and Abdelnabi, Sahar and Mishra, Shailesh and Endres, Christoph and Holz, Thorsten and Fritz, Mario},
title = {Not What You've Signed Up For: Compromising Real-World LLM-Integrated Applications with Indirect Prompt Injection},
year = {2023},
isbn = {9798400702600},
publisher = {Association for Computing Machinery},
address = {New York, NY, USA},
url = {https://doi.org/10.1145/3605764.3623985},
doi = {10.1145/3605764.3623985},
booktitle = {Proceedings of the 16th ACM Workshop on Artificial Intelligence and Security},
pages = {79–90},
numpages = {12},
location = {Copenhagen, Denmark},
series = {AISec '23}
}

@inproceedings{ji2023towardsmitigatingllmhallucination,
    title = "Towards Mitigating {LLM} Hallucination via Self Reflection",
    author = "Ji, Ziwei  and
      Yu, Tiezheng  and
      Xu, Yan  and
      Lee, Nayeon  and
      Ishii, Etsuko  and
      Fung, Pascale",
    editor = "Bouamor, Houda  and
      Pino, Juan  and
      Bali, Kalika",
    booktitle = "Findings of the Association for Computational Linguistics: EMNLP 2023",
    month = dec,
    year = "2023",
    address = "Singapore",
    publisher = "Association for Computational Linguistics",
    url = "https://aclanthology.org/2023.findings-emnlp.123/",
    doi = "10.18653/v1/2023.findings-emnlp.123",
    pages = "1827--1843"
}

@misc{kumar2024manipulatinglargelanguagemodels,
      title={Manipulating Large Language Models to Increase Product Visibility}, 
      author={Aounon Kumar and Himabindu Lakkaraju},
      year={2024},
      eprint={2404.07981},
      archivePrefix={arXiv},
      primaryClass={cs.IR},
      url={https://arxiv.org/abs/2404.07981}, 
}

@ARTICLE{kowalczyk2024EnhancingSEOinsingle-page,
  author={Kowalczyk, Karolina and Szandala, Tomasz},
  journal={IEEE Access}, 
  title={Enhancing SEO in Single-Page Web Applications in Contrast With Multi-Page Applications}, 
  year={2024},
  volume={12},
  number={},
  pages={11597-11614},
  doi={10.1109/ACCESS.2024.3355740}}

@INPROCEEDINGS{kanara2024pythondrivenkeywordanalysis,
  author={Kanara, Aditya Pranav and Kumari, Priya and Prathap, Boppuru Rudra},
  booktitle={2024 10th International Conference on Advanced Computing and Communication Systems (ICACCS)}, 
  title={Python Driven Keyword Analysis for SEO Optimization}, 
  year={2024},
  volume={1},
  number={},
  pages={1170-1176},
  doi={10.1109/ICACCS60874.2024.10717132}}

@inproceedings{li2024learningtorewriteprompts,
author = {Li, Cheng and Zhang, Mingyang and Mei, Qiaozhu and Kong, Weize and Bendersky, Michael},
title = {Learning to Rewrite Prompts for Personalized Text Generation},
year = {2024},
isbn = {9798400701719},
publisher = {Association for Computing Machinery},
address = {New York, NY, USA},
url = {https://doi.org/10.1145/3589334.3645408},
doi = {10.1145/3589334.3645408},
booktitle = {Proceedings of the ACM Web Conference 2024},
pages = {3367–3378},
numpages = {12},
location = {Singapore, Singapore},
series = {WWW '24}
}

@misc{li2025deepthinkaligninglanguagemodels,
      title={DeepThink: Aligning Language Models with Domain-Specific User Intents}, 
      author={Yang Li and Mingxuan Luo and Yeyun Gong and Chen Lin and Jian Jiao and Yi Liu and Kaili Huang},
      year={2025},
      eprint={2502.05497},
      archivePrefix={arXiv},
      primaryClass={cs.CL},
      url={https://arxiv.org/abs/2502.05497}, 
}

@inproceedings{liuetal2023gevalnlgevalutationusing,
    title = "{G}-Eval: {NLG} Evaluation using Gpt-4 with Better Human Alignment",
    author = "Liu, Yang  and
      Iter, Dan  and
      Xu, Yichong  and
      Wang, Shuohang  and
      Xu, Ruochen  and
      Zhu, Chenguang",
    editor = "Bouamor, Houda  and
      Pino, Juan  and
      Bali, Kalika",
    booktitle = "Proceedings of the 2023 Conference on Empirical Methods in Natural Language Processing",
    month = dec,
    year = "2023",
    address = "Singapore",
    publisher = "Association for Computational Linguistics",
    url = "https://aclanthology.org/2023.emnlp-main.153/",
    doi = "10.18653/v1/2023.emnlp-main.153",
    pages = "2511--2522"
}

@inproceedings{maoetal2023largelanguagemodelsknow,
    title = "Large Language Models Know Your Contextual Search Intent: A Prompting Framework for Conversational Search",
    author = "Mao, Kelong  and
      Dou, Zhicheng  and
      Mo, Fengran  and
      Hou, Jiewen  and
      Chen, Haonan  and
      Qian, Hongjin",
    editor = "Bouamor, Houda  and
      Pino, Juan  and
      Bali, Kalika",
    booktitle = "Findings of the Association for Computational Linguistics: EMNLP 2023",
    month = dec,
    year = "2023",
    address = "Singapore",
    publisher = "Association for Computational Linguistics",
    url = "https://aclanthology.org/2023.findings-emnlp.86/",
    doi = "10.18653/v1/2023.findings-emnlp.86",
    pages = "1211--1225"
}

@inproceedings{pfrommer2024rankingmanipulationforconversational,
    title = "Ranking Manipulation for Conversational Search Engines",
    author = "Pfrommer, Samuel  and
      Bai, Yatong  and
      Gautam, Tanmay  and
      Sojoudi, Somayeh",
    editor = "Al-Onaizan, Yaser  and
      Bansal, Mohit  and
      Chen, Yun-Nung",
    booktitle = "Proceedings of the 2024 Conference on Empirical Methods in Natural Language Processing",
    month = nov,
    year = "2024",
    address = "Miami, Florida, USA",
    publisher = "Association for Computational Linguistics",
    url = "https://aclanthology.org/2024.emnlp-main.534/",
    doi = "10.18653/v1/2024.emnlp-main.534",
    pages = "9523--9552"
}

@article{nagpal2021keywordselectionstrategiesin,
title = {Keyword Selection Strategies in Search Engine Optimization: How Relevant is Relevance?},
journal = {Journal of Retailing},
volume = {97},
number = {4},
pages = {746-763},
year = {2021},
note = {SI: Metrics and Analytics},
issn = {0022-4359},
doi = {https://doi.org/10.1016/j.jretai.2020.12.002},
url = {https://www.sciencedirect.com/science/article/pii/S0022435920300944},
author = {Mayank Nagpal and J. Andrew Petersen}
}

@misc{sarkar2025conversationaluseraiinterventionstudy,
      title={Conversational User-AI Intervention: A Study on Prompt Rewriting for Improved LLM Response Generation}, 
      author={Rupak Sarkar and Bahareh Sarrafzadeh and Nirupama Chandrasekaran and Nagu Rangan and Philip Resnik and Longqi Yang and Sujay Kumar Jauhar},
      year={2025},
      eprint={2503.16789},
      archivePrefix={arXiv},
      primaryClass={cs.CL},
      url={https://arxiv.org/abs/2503.16789}, 
}

@article{shahzad2020thenewtrendfor,
  title={The new trend for search engine optimization, tools and techniques},
  author={Shahzad, Asim and Jacob, Deden Witarsyah and Nawi, Nazri Mohd and Mahdin, Hairulnizam and Saputri, Marheni Eka},
  journal={Indonesian Journal of Electrical Engineering and Computer Science},
  volume={18},
  number={3},
  pages={1568--1583},
  year={2020}
}

@inproceedings{shi2024optimizationbasedpromptinjection,
author = {Shi, Jiawen and Yuan, Zenghui and Liu, Yinuo and Huang, Yue and Zhou, Pan and Sun, Lichao and Gong, Neil Zhenqiang},
title = {Optimization-based Prompt Injection Attack to LLM-as-a-Judge},
year = {2024},
isbn = {9798400706363},
publisher = {Association for Computing Machinery},
address = {New York, NY, USA},
url = {https://doi.org/10.1145/3658644.3690291},
doi = {10.1145/3658644.3690291},
booktitle = {Proceedings of the 2024 on ACM SIGSAC Conference on Computer and Communications Security},
pages = {660–674},
numpages = {15},
location = {Salt Lake City, UT, USA},
series = {CCS '24}
}

@inproceedings{shinn2023reflexionlanguageagentswith,
 author = {Shinn, Noah and Cassano, Federico and Gopinath, Ashwin and Narasimhan, Karthik and Yao, Shunyu},
 booktitle = {Advances in Neural Information Processing Systems},
 doi = {10.52202/075280-0377},
 editor = {A. Oh and T. Naumann and A. Globerson and K. Saenko and M. Hardt and S. Levine},
 pages = {8634--8652},
 publisher = {Curran Associates, Inc.},
 title = {Reflexion: language agents with verbal reinforcement learning},
 url = {https://proceedings.neurips.cc/paper_files/paper/2023/file/1b44b878bb782e6954cd888628510e90-Paper-Conference.pdf},
 volume = {36},
 year = {2023}
}

@INPROCEEDINGS{samarah2024utilizingllmsforenhancing,
  author={Samarah, Tariq and Alrawashdeh, Tawfiq and Mughaid, Ala’ S and AlZu’Bi, Shadi},
  booktitle={2024 2nd International Conference on Foundation and Large Language Models (FLLM)}, 
  title={Utilizing LLMs for Enhancing Search Engine Optimization Strategies in Digital Marketing}, 
  year={2024},
  volume={},
  number={},
  pages={284-288},
  doi={10.1109/FLLM63129.2024.10852433}}

@inproceedings{shu2024rewritelmaninstruction-tuned,
  title={Rewritelm: An instruction-tuned large language model for text rewriting},
  author={Shu, Lei and Luo, Liangchen and Hoskere, Jayakumar and Zhu, Yun and Liu, Yinxiao and Tong, Simon and Chen, Jindong and Meng, Lei},
  booktitle={Proceedings of the AAAI Conference on Artificial Intelligence},
  volume={38},
  number={17},
  pages={18970--18980},
  year={2024}
}

@article {Ward2017WhywhosewhatandhowAframeworkforknowledgemobilisers,
      author = "Vicky Ward",
      title = "Why, whose, what and how? A framework for knowledge mobilisers",
      journal = "Evidence and Policy",
      year = "2017",
      publisher = "Policy Press",
      address = "Bristol, UK",
      volume = "13",
      number = "3",
      doi = "10.1332/174426416X14634763278725",
      pages=      "477 - 497",
      url = "https://bristoluniversitypressdigital.com/view/journals/evp/13/3/article-p477.xml"
}

@inproceedings{yanetal2024mirrormultipleperspectiveselfreflectionmethod,
    title = "Mirror: Multiple-perspective Self-Reflection Method for Knowledge-rich Reasoning",
    author = "Yan, Hanqi  and
      Zhu, Qinglin  and
      Wang, Xinyu  and
      Gui, Lin  and
      He, Yulan",
    editor = "Ku, Lun-Wei  and
      Martins, Andre  and
      Srikumar, Vivek",
    booktitle = "Proceedings of the 62nd Annual Meeting of the Association for Computational Linguistics (Volume 1: Long Papers)",
    month = aug,
    year = "2024",
    address = "Bangkok, Thailand",
    publisher = "Association for Computational Linguistics",
    url = "https://aclanthology.org/2024.acl-long.382/",
    doi = "10.18653/v1/2024.acl-long.382",
    pages = "7086--7103"
}

@inproceedings{zhangetal2024selfcontrastbetterreflectionthrough,
    title = "Self-Contrast: Better Reflection Through Inconsistent Solving Perspectives",
    author = "Zhang, Wenqi  and
      Shen, Yongliang  and
      Wu, Linjuan  and
      Peng, Qiuying  and
      Wang, Jun  and
      Zhuang, Yueting  and
      Lu, Weiming",
    editor = "Ku, Lun-Wei  and
      Martins, Andre  and
      Srikumar, Vivek",
    booktitle = "Proceedings of the 62nd Annual Meeting of the Association for Computational Linguistics (Volume 1: Long Papers)",
    month = aug,
    year = "2024",
    address = "Bangkok, Thailand",
    publisher = "Association for Computational Linguistics",
    url = "https://aclanthology.org/2024.acl-long.197/",
    doi = "10.18653/v1/2024.acl-long.197",
    pages = "3602--3622"
}

@article{Škėrienė2020problemsolvingthroughvalues,
title = {Problem solving through values: A challenge for thinking and capability development},
journal = {Thinking Skills and Creativity},
volume = {37},
pages = {100694},
year = {2020},
issn = {1871-1871},
doi = {https://doi.org/10.1016/j.tsc.2020.100694},
url = {https://www.sciencedirect.com/science/article/pii/S1871187120301681},
author = {Sandrita Škėrienė and Palmira Jucevičienė}
}

@inproceedings{nestaas2025adversarialsearchengineoptimization,
 author = {Nestaas, Fredrik and Debenedetti, Edoardo and Tramer, Florian},
 booktitle = {International Conference on Learning Representations},
 editor = {Y. Yue and A. Garg and N. Peng and F. Sha and R. Yu},
 pages = {4857--4888},
 title = {Adversarial Search Engine Optimization for Large Language Models},
 url = {https://proceedings.iclr.cc/paper_files/paper/2025/file/0f12b3c36a781120c4f60e90e855868d-Paper-Conference.pdf},
 volume = {2025},
 year = {2025}
}

@misc{bagga2026egeotestbedgenerativeengine,
      title={E-GEO: A Testbed for Generative Engine Optimization in E-Commerce}, 
      author={Puneet S. Bagga and Vivek F. Farias and Tamar Korkotashvili and Tianyi Peng and Yuhang Wu},
      year={2026},
      eprint={2511.20867},
      archivePrefix={arXiv},
      primaryClass={cs.IR},
      url={https://arxiv.org/abs/2511.20867}, 
}

@inproceedings{bardas2025automaticdocumenteditingfor,
author = {Bardas, Niv and Mordo, Tommy and Kurland, Oren and Tennenholtz, Moshe},
title = {Automatic Document Editing for Improved Ranking},
year = {2025},
isbn = {9798400715921},
publisher = {Association for Computing Machinery},
address = {New York, NY, USA},
url = {https://doi.org/10.1145/3726302.3730168},
doi = {10.1145/3726302.3730168},
booktitle = {Proceedings of the 48th International ACM SIGIR Conference on Research and Development in Information Retrieval},
pages = {2779–2783},
numpages = {5},
location = {Padua, Italy},
series = {SIGIR '25}
}

@book{lewandowski2023understandingearchengines,
  title={Understanding search engines},
  author={Lewandowski, Dirk},
  year={2023},
  publisher={Springer}
}

@misc{ramakrishna2026queryimpliedgenerativeengine,
      title={Query Implied Generative Engine Optimization}, 
      author={Shilpa Ramakrishna and William B. Andreopoulos},
      year={2026},
      eprint={2609.27845},
      archivePrefix={arXiv},
      primaryClass={cs.IR},
      url={https://arxiv.org/abs/2609.27845}, 
}

@article{iqbal2022searchengineoptimizationseo,
  title={Search engine optimization (SEO): A study of important key factors in achieving a better search engine result page (SERP) position},
  author={Iqbal, Muhammad and Khalid, Muhammad Noman and Manzoor, Amir and Abid, Malik Munib and Shaikh, Nazir Ahmed},
  journal={Sukkur IBA Journal of Computing and Mathematical Sciences},
  volume={6},
  number={1},
  pages={1--15},
  year={2022}
}

@misc{chen2025ccgseobenchcontentcentricbenchmarkmeasuring,
      title={CC-GSEO-Bench: A Content-Centric Benchmark for Measuring Source Influence in Generative Search Engines}, 
      author={Qiyuan Chen and Jiahe Chen and Hongsen Huang and Qian Shao and Jintai Chen and Renjie Hua and Hongxia Xu and Ruijia Wu and Ren Chuan and Jian Wu},
      year={2025},
      eprint={2509.05607},
      archivePrefix={arXiv},
      primaryClass={cs.CL},
      url={https://arxiv.org/abs/2509.05607}, 
}

\appendix

\section{Core Prompts Used in RAID G-SEO}
\label{app:Core Prompts Used in RAID G-SEO}
RAID's core prompts comprise an opening description, task instructions, and input and output formats. The opening describes the information provided to the model, task instructions specify the operations, and input and output formats define the required fields and expected results. Table~\ref{tab:Structured_overview_of_RAID_s_core_prompts} organizes the task instructions around RAID's four-stage intent-driven optimization framework with 4W multi-role deep reflection embedded in the second stage of intent inference and refinement.

\begin{table}[!b]
\centering
\caption{Structured overview of RAID's core prompts, including opening descriptions, task instructions covering the four-stage optimization framework and 4W multi-role deep reflection, and input/output formats.}
\label{tab:Structured_overview_of_RAID_s_core_prompts}
\footnotesize
\renewcommand{\arraystretch}{1.1}

\begin{tabularx}{\columnwidth}{@{}>{\raggedright\arraybackslash}X@{}}
\toprule
\textbf{Core Prompts of RAID G-SEO}\\
\midrule

\textbf{Opening}
\par\smallskip
You will be given a source of web content, along with its summary, primary search intent, and refined information needs when applicable.
\\

\midrule
\textbf{Task Instructions}
\par\medskip

\textbf{Step 1: Content Summarization}
\par
Summarize the source clearly and concisely by extracting the key information and main points, avoiding irrelevant details, to make it understandable for readers who have not seen the original.

\par\medskip
\textbf{Step 2: Intent Inference}
\par
Analyze the source and its summary to identify the primary search intent, describing what a reader would want to learn from the content.

\par\medskip
\textbf{4W Multi-Role Deep Reflection (Step 2)}
\par
Reflect on the primary intent from 1--3 likely reader roles:
\par
\hspace*{1em}\textbullet\ \textbf{WHO}: Identify the reader role.
\par
\hspace*{1em}\textbullet\ \textbf{WHAT}: State the information this role would seek from the content.
\par
\hspace*{1em}\textbullet\ \textbf{WHY}: Identify what the primary intent leaves implicit for this role, or confirm its sufficiency.
\par
\hspace*{1em}\textbullet\ \textbf{HOW}: Refine the intent to make that information need explicit.
\par
Integrate the primary need and useful role-specific distinctions into a coherent refined intent of 1--3 sentences. Limit the needs to information provided by the source. If refinement adds nothing useful, retain the primary intent.

\par\medskip
\textbf{Step 3: Step Planning}
\par
Based on the primary and refined intents, determine the best optimization strategy from multiple angles to increase visibility and user engagement. (For each supported need, identify relevant information already in the original source and make it easier to locate, understand, and cite. Treat refined needs as complementary coverage requirements rather than minor adjustments.)

\par\medskip
\textbf{Step 4: Content Rewriting}
\par
Follow the above optimization steps to rewrite the source, preserving its core meaning and key information.

\par\medskip
\textbf{Common Output Constraint}
\par
Output only the result required by the current task, without additional explanations or analysis. 
\\

\midrule
\textbf{Input Format}
\par\smallskip
Original Source,
Source Summary,
Primary Intent,
Refined Intent
\\

\midrule
\textbf{Output Format}
\par\smallskip
Summary: [Your summary]
\par
Primary Intent: [Main information need]
\par
Refined Intent: [Primary and complementary information needs]
\par
Optimized Source: [Your optimized source]
\\

\bottomrule
\end{tabularx}
\end{table}

\section{Query Variant Generation and Validation}
\label{app:Query Variant Generation and Validation}
To simulate diverse potential information needs associated with the same topic, we use GPT-4 to initially generate four topic-related variants for each of the 1,000 original queries in the GEO-BENCH test split. The generation stage takes only the original query as input and encourages exploration of the topic from different perspectives, including related concepts, practical applications, and contrasting viewpoints. The complete prompt used in our implementation is presented in Table~\ref{tab:Prompt_for_generating_four_topic-related_query_variants_from_an_original_GEO-BENCH_query_to_explore_diverse_potential_information_needs}.
To improve the quality of the expanded dataset, we conducted multiple rounds of validation and revision, focusing on two aspects. First, we examined the topical relevance of each query variant to its original query. Second, we assessed whether the fixed set of five content sources associated with the original query provided substantive support for the variant's core information needs. Throughout this process, the original queries and their associated content sources remained unchanged. Among the 4,000 initially generated variants, 1,128 were revised. The final quality audit found that 3,988 variants (99.70\%) satisfied the quality requirements, while the remaining 12 variants were associated with three pre-existing mismatches between original queries and their corresponding source sets. Following manual screening, we excluded these three anomalous groups, retaining 997 valid extended samples.

\begin{table}[t]
\centering
\caption{Prompt for generating four topic-related query variants from an original GEO-BENCH query to explore diverse potential information needs.}
\label{tab:Prompt_for_generating_four_topic-related_query_variants_from_an_original_GEO-BENCH_query_to_explore_diverse_potential_information_needs}
\footnotesize
\renewcommand{\arraystretch}{1.15}

\begin{tabularx}{\columnwidth}{@{}>{\raggedright\arraybackslash}X@{}}
\toprule

\textbf{Query Variant Generation Prompt} \\

\midrule

You are a user of the Generative Search Engine, seeking to explore a specific topic. You will be given a query representing what you want to learn about. Your task is to generate 4 new queries that approach the topic from different perspectives. Ensure that:

\par\smallskip
1. Each query is relevant to the original but not a direct repetition.

\par\smallskip
2. The queries cover a range of perspectives, such as related concepts, practical applications, contrasting viewpoints, emerging trends, or any other meaningful dimension. But do not limit to these directions.

\par\smallskip
3. Output ONLY 1 query per line.

\par\medskip
\textbf{\#\#\# Input Format}

\par
\texttt{\{query\_original\}}

\par\medskip
\textbf{\#\#\# Output Format}

\par
\texttt{query\_1}
\par
\texttt{query\_2}
\par
\texttt{query\_3}
\par
\texttt{query\_4}

\\
\bottomrule
\end{tabularx}
\end{table}

\begin{table}[t]
\centering

\caption{Prompt for generating scoring criteria and evaluation steps across seven subjective impression dimensions. Given each dimension's description, it refines the criteria and specifies steps for evaluating the target source, including a zero-score rule for uncited sources.}
\label{tab:Prompt_for_generating_scoring_criteria_and_evaluation_steps_across_seven_subjective_impression_dimensions}
\footnotesize
\renewcommand{\arraystretch}{1.15}
\setlength{\tabcolsep}{6pt}

\begin{tabularx}{\columnwidth}{@{}>{\raggedright\arraybackslash}X@{}}
\toprule
\textbf{G-EVAL Prompt Generation Prompt} \\
\midrule

You are an evaluator for answers generated by the Generative Search Engine, providing subjective evaluations...

\par\medskip
Your tasks:

\par\smallskip
1. Refine the criteria: Review the given description and improve it to be concise, consistency, and completeness.

\par\smallskip
2. Propose an Evaluation Procedure: Based on the refined criteria, outline a clear, concise, step-by-step process specifically for scoring Source [1] only. If the source not exsist, score should be 0.

\par\smallskip
Please ensure your generation is easy to understand.

\par\medskip
\#\#\# Input Format

\par\smallskip
\textbf{Example:} Relevance of Citation to Query (0-5): The dimension evaluates the degree to which the citation from Source [1] is related to the query, indicating how useful and pertinent the information provided to address the query....

\par\medskip
\#\#\# Output Format

\par\smallskip
Evaluation Criteria: [...]

\par\smallskip
Evaluation Steps:

\par\smallskip
1. Step 1 description...

\par\smallskip
2. Step 2 description...

\par\smallskip
...

\\
\bottomrule
\end{tabularx}
\end{table}

\begin{table}[t]
\centering
\caption{Unified G-Eval prompt template for evaluating seven subjective impression dimensions. It incorporates dimension-specific scoring criteria and evaluation steps, takes the user query and generated response as inputs, and requires only the final score for the target source.}
\label{tab:Unified_G-EVAL_prompt_template_for_evaluating_seven_subjective_impression_dimensions}
\footnotesize
\renewcommand{\arraystretch}{1.15}
\setlength{\tabcolsep}{6pt}

\begin{tabularx}{\columnwidth}{@{}>{\raggedright\arraybackslash}X@{}}
\toprule
\textbf{Unified Evaluation Procedure Template for G-eval} \\
\midrule

You are a user of Generative Search Engine, capable of providing subjective evaluations of retrieved answers... Please read and understand these instructions carefully, and then proceed with your scoring (score ONLY).

\par\medskip
\#\#\# Evaluation Criteria

\par\smallskip
[Evaluation Criteria]: [...]

\par\medskip
\#\#\# Evaluation Steps

\par\smallskip
[...]

\par\medskip
\#\#\# Input Format

\par\smallskip
Query: \{query\}

\par\smallskip
Answer: \{answer\}

\par\medskip
\#\#\# Output Format (score ONLY)

\par\smallskip
[score]

\\
\bottomrule
\end{tabularx}
\end{table}

\section{G-Eval Prompt Generation and Evaluation Template}
\label{app:G-eval Prompt Generation Prompt and Evaluation Prompt Template}
G-Eval \cite{liuetal2023gevalnlgevalutationusing} combines chain-of-thought (CoT) reasoning with structured scoring. GEO \cite{aggarwal2024geogenerativeengineoptimization} adopts this framework to assess subjective source visibility in GSE responses. We retain G-Eval's stepwise evaluation approach while adapting GEO's prompts for seven subjective impression dimensions. Specifically, GPT-4o refines dimension-specific criteria and generates evaluation steps through a prompt-generate-prompt procedure, as detailed in Table~\ref{tab:Prompt_for_generating_scoring_criteria_and_evaluation_steps_across_seven_subjective_impression_dimensions}. The resulting criteria and steps are incorporated into the unified template in Table~\ref{tab:Unified_G-EVAL_prompt_template_for_evaluating_seven_subjective_impression_dimensions}. Given a user query and generated response, GLM-4-9B-0414 uses this template to assign a 0--5 score to the target source for the specified dimension and outputs only the final score. Uncited sources score 0.

\begin{table*}[!t]
\centering
\caption{RAID case study on a gift-recommendation source from the expanded GEO-BENCH (one original query and four topic-related variants, all unseen during optimization, with target source [3]). (a) Original and rewritten source excerpts with implemented edits. (b) Before/after GSE response excerpts, response changes, and objective Overall (O) and subjective Average (S) RIS. Citation counts refer to the full responses of the illustrated generation, while RIS aggregates five generations per query.}
\label{tab:RAID_case_study_on_an_SOP_source_from_the_expanded_GEO-BENCH}
\begingroup
\setlength{\tabcolsep}{2.75pt}
\renewcommand{\arraystretch}{1.0}
\scriptsize
\begin{tabular}{@{}p{0.127\textwidth}p{0.248\textwidth}p{0.248\textwidth}p{0.244\textwidth}@{\hspace{7pt}}l@{}}
\toprule
\multicolumn{5}{c}{\textbf{(a) Source Revision}} \\
\midrule
\multicolumn{5}{@{}c@{}}{%
\begin{tabular}{@{}p{0.468\textwidth}@{\hspace{10pt}}p{0.468\textwidth}@{}}
\multicolumn{1}{c}{\textbf{Original Source}} & \multicolumn{1}{c}{\textbf{Rewritten Source}} \\
\midrule
\textit{Gifting can be difficult --- whether it's during the holiday season or any other time of year. Finding gifts for the people we know best can seem impossible, much less finding a gift for someone we don't know. \ldots{}} & \textit{\ldots{} \textbf{Navigating the Art of Gifting: Thoughtful Ideas for Any Occasion} \ldots{} This guide provides curated suggestions for inexpensive, sympathy, celebratory gifts, and gift baskets, emphasizing the importance of kindness and thoughtfulness in every gesture. \ldots{}} \\
\addlinespace[1pt]
\textit{\ldots{} Our Top Picks \ldots{} Inexpensive Gifts \ldots{} Sympathy Gifts \ldots{} Celebratory Gifts \ldots{} Gift Baskets \ldots{}} & \textit{\ldots{} \textbf{Our Top Picks by Category:} \ldots{} \textbf{Inexpensive Gifts for Strangers} \ldots{} \textbf{Sympathy Gifts for Strangers} \ldots{} \textbf{Celebratory Gifts for Strangers} \ldots{} \textbf{Gift Baskets for Strangers} \ldots{}} \\
\addlinespace[1pt]
\textit{\ldots{} 1. A nice journal \ldots{} Journaling may not be habitual for everyone or even something your friend is interested in right now. However, it's never too late to begin, and the journal you gift may be the motivation they need to do so. \ldots{}} & \textit{\ldots{} \textbf{A Nice Journal:} Journaling isn't a prerequisite for everyone, but a well-chosen journal, perhaps one with prompts, could inspire someone new to this practice. Consider one that complements their interests or style. \ldots{}} \\
\addlinespace[1pt]
\textit{\ldots{} 7. A donation on their behalf \ldots{} Making a donation to a cause that's special to them or their late loved one helps pay kindness and generosity forward. \ldots{}} & \textit{\ldots{} \textbf{A donation on their behalf:} For those less inclined towards physical gifts or during sensitive times like grief, making a donation to a cause they care about is a meaningful way to show support. \ldots{}} \\
\addlinespace[1pt]
\textit{\ldots{} But you shouldn't give up. After all, it's your kindness and the gesture itself that counts. There are plenty of gift ideas out there that are always pleasing to just about any crowd. \ldots{}} & \textit{\ldots{} Ultimately, gifting is about the thought and kindness behind the gesture. Even the simplest present, given with care, can make a meaningful impact.} \\
\end{tabular}} \\
\midrule
\multicolumn{5}{c}{\textbf{Implemented Optimization Operations}} \\
\midrule

\multicolumn{5}{@{}p{0.957\textwidth}@{}}{\textbf{E1:} Add a descriptive title and introductory overview. \textbf{E2:} Make existing gift-category headings explicitly address strangers. \textbf{E3:} Recast gift suggestions as concise labeled entries. \textbf{E4:} Revise explanations of journals, donations, and other recommendations. \textbf{E5:} Foreground the thoughtfulness of the gesture in the conclusion. \textbf{E6:} Compress the source from 1,437 to 875 whitespace-delimited tokens.} \\
\midrule
\multicolumn{5}{c}{\textbf{(b) Query-Specific Responses}} \\
\midrule
\multicolumn{1}{c}{\textbf{Query}} & \multicolumn{1}{c}{\textbf{Before-optimization Response}} & \multicolumn{1}{c}{\textbf{After-optimization Response}} & \multicolumn{1}{c}{\textbf{Response Change}} & \multicolumn{1}{c}{\textbf{RIS}} \\
\midrule

\textbf{Q0}\quad What are the best birthday gifts for an acquaintance that I don't know very well? & \ldots{} Inexpensive yet thoughtful options include a leather journal for \$37.95, handmade lavender soy candle for \$19.95, and an art book for \$13.89, which can inspire creativity or relaxation \textbf{[3]}. \ldots{} & \ldots{} gift cards to local restaurants, comedy clubs, or wellness services \ldots{} [4]. \ldots{} for someone you don\'t know well [1][2]\textbf{[3]}[4][5]. \ldots{} & Priced gifts become neutral options. [3] citations drop from 5 to 1, shifting from exclusive gift claims to a shared final citation. & \begin{tabular}[t]{@{}l@{}}\textbf{O:} $-2.57$ \\ \textbf{S:} $+23.15$\end{tabular} \\
\addlinespace[1pt]
\textbf{Q1}\quad What are some safe and neutral birthday gifts for acquaintances? & \ldots{} These options strike a balance between being safe, neutral, and thoughtful, making them suitable for acquaintances while still showing consideration and kindness [1][2]\textbf{[3]}[4][5]. \ldots{} & \ldots{} while gift cards to local eateries, fitness classes, or bookstores are practical and neutral [2]\textbf{[3]}. \ldots{} These gifts balance usefulness, neutrality, and affordability, making them suitable for someone you don't know well [2]\textbf{[3]}. \ldots{} & Gift lists become concise categories. [3] citations rise from 1 to 2, appearing in gift-card advice and the conclusion. & \begin{tabular}[t]{@{}l@{}}\textbf{O:} $+3.21$ \\ \textbf{S:} $+12.71$\end{tabular} \\
\addlinespace[1pt]
\textbf{Q2}\quad What are the most popular birthday gifts for people you don't know well? & \ldots{} These gifts strike a balance between being considerate and respectful, ensuring the recipient feels appreciated without the giver needing to have extensive knowledge of their preferences [1][2]\textbf{[3]}[4][5]. & \ldots{} a nice journal, a candle, a coloring book, and lottery scratch-offs \textbf{[3]}. \ldots{} neutral, useful, and fun, often within a reasonable price point [2]. \ldots{} & General advice shifts to inexpensive gifts. [3] stays at one citation, moving from a shared conclusion to exclusive gift-list support. & \begin{tabular}[t]{@{}l@{}}\textbf{O:} $-1.77$ \\ \textbf{S:} $+5.52$\end{tabular} \\
\addlinespace[1pt]
\textbf{Q3}\quad What are some unique but universally appealing birthday gifts for an acquaintance? & \ldots{} Practical and neutral choices, like a multipurpose pot, game to break the ice, or zen garden, are crowd-pleasers under \$25 [2]. \ldots{} useful, fun, and noncommittal \ldots{} [2][5]. & \ldots{} Practical and neutral items, like multipurpose plant pots, game sets, or reusable straws, are also crowd-pleasers that cater to a wide range of interests [2]\textbf{[3]}. \ldots{} ensuring the gesture is thoughtful and respectful [2]\textbf{[3]}[4][5]. \ldots{} & Gift advice adds reusable items. [3] citations rise from 0 to 2 in recommendations and the final claim, alongside other sources. & \begin{tabular}[t]{@{}l@{}}\textbf{O:} $+5.42$ \\ \textbf{S:} $+12.06$\end{tabular} \\
\addlinespace[1pt]
\textbf{Q4}\quad What are some thoughtful birthday gifts that don't require personal knowledge of the recipient? & \ldots{} Thoughtful birthday gifts that don't require personal knowledge of the recipient can include subscription services, gift cards, and neutral items suitable for various tastes [1][2]\textbf{[3]}[4]. \ldots{} & \ldots{} Personalized items like monogrammed items, virtual messages, or customized gifts can add a special touch without needing detailed knowledge of the recipient's preferences [1][2]\textbf{[3]}[4]. & Personalized gifts become prominent. [3] citations rise from 2 to 4 across the opening and recommendations, with no exclusive citations. & \begin{tabular}[t]{@{}l@{}}\textbf{O:} $+7.04$ \\ \textbf{S:} $-1.35$\end{tabular} \\
\bottomrule
\end{tabular}
\endgroup
\end{table*}

\section{Experiment Details}
\subsection{Software and Hardware Environment}
\label{app:Software and Hardware Environment}
The local experimental environment was an Ubuntu 20.04 server with two NVIDIA GeForce RTX 3090 GPUs, each with 24 GB of VRAM. We used Python 3.9.19, PyTorch 2.1.0 with CUDA 11.8 support, and Hugging Face Transformers 4.51.3, with dependencies managed by Conda. We loaded and ran local models using Transformers and accessed GPT models through the OpenAI API.

\begin{table}[H]
\centering
\caption{Core configurations of LLMs used in this work, including task-specific applications and generation settings.}
\label{tab:Core_configurations_of_LLMs_used_in_this_work}

\footnotesize
\renewcommand{\arraystretch}{1.15}
\setlength{\tabcolsep}{4pt}

\begin{threeparttable}
\begin{tabularx}{\columnwidth}{@{}
>{\raggedright\arraybackslash}p{0.23\columnwidth}
>{\raggedright\arraybackslash}X@{}}
\toprule

\textbf{Model} & \textbf{Application and Generation Settings} \\

\midrule

GPT-4
& \textbf{Query Variants Generation}
\par
Temp.: 0.5; Top-$p$: 1
\par
Sampling: --; Max. tokens: 1024 \\

\midrule

GPT-4o
& \textbf{G-Eval Prompt Generation}
\par
Temp.: 0.5; Top-$p$: 1
\par
Sampling: --; Max. tokens: 3192 \\

\midrule

GLM-4-9B-0414
& \textbf{GSE Response Generation Simulation}
\par
Temp.: 0.5; Top-$p$: 1
\par
Sampling: True; Max. tokens: 1024 \\

\addlinespace[0pt]

& \textbf{G-SEO Methods Implementation}
\par
Temp.: --; Top-$p$: 1
\par
Sampling: False; Max. tokens: 3192 \\

\addlinespace[0pt]

& \textbf{G-Eval Subjective Impression Evaluation}
\par
Temp.: --; Top-$p$: 1
\par
Sampling: False; Max. tokens: 2 \\
\bottomrule
\end{tabularx}
\end{threeparttable}
\end{table}

\subsection{Configuration Parameters of LLMs}
\label{app:Configuration Parameters of LLMs}
We report task-specific LLM configurations for dataset construction and experimental evaluation in Table~\ref{tab:Core_configurations_of_LLMs_used_in_this_work}, including applications, temperature, top-$p$, sampling settings, and maximum output tokens.

\section{A Case of RAID G-SEO}
\label{app:A Case of RAID G-SEO}
To further examine how a single content rewrite performs across unseen queries, we analyze the revisions to a gift-recommendation source and its use in GSE responses. RAID rewrites only the target source [3], leaving the other four sources unchanged. Table~\ref{tab:RAID_case_study_on_an_SOP_source_from_the_expanded_GEO-BENCH} compares the implemented edits with before- and after-optimization response excerpts across five unseen queries, showing how target-source citations and generated responses vary across queries.

\section{Limitations}
RAID uses multi-role intent refinement to address diverse potential information needs, yet its visibility gains are not uniform across unseen queries. Our evaluation uses fixed source sets and single-turn responses generated by GLM-4-9B-0414 in a simulated GSE, measuring visibility after source inclusion rather than end-to-end retrieval performance. The same model is also used for content optimization and subjective scoring, which may introduce model-specific evaluation effects. Further validation with independent evaluators, dynamic retrieval, and deployed GSEs could help assess the generalizability of these findings.

\section{Ethics Statement}
We used LLMs in accordance with applicable licenses and terms of service. We conducted controlled experiments in a simulated GSE environment without intervening in or modifying any deployed search services. We used LLMs to generate semantically related query variants to extend GEO-BENCH and rewrote selected content sources to evaluate the effects of content optimization. The original queries and their variants were used for benchmark construction and evaluation, without being provided to the content optimizer and the rewritten content was used only in controlled experiments. These generated or modified materials served solely as experimental data and should not be regarded as independently fact-checked information or as representing the authors' views or positions. We acknowledge that G-SEO techniques could be misused to amplify misleading information or unfairly influence source visibility in generative search.

\end{document}